\newcommand{\tsecompldate}{18th November 2016}

\documentclass[a4paper,12pt]{article}
\usepackage{natbib}

\usepackage{amsmath,amssymb,amscd}

\usepackage{graphicx}

\usepackage{hyperref}

\providecommand{\citep}[1]{\cite{#1}}
\newcommand{\bea}{\begin{eqnarray}}
\newcommand{\eea}{\end{eqnarray}}
\newcommand{\beq}{\begin{equation}}
\newcommand{\eeq}{\end{equation}}

\newcommand{\tref}[1]{(\ref{#1})}
\newcommand{\figref}[1]{Fig.~\ref{#1}}
\newcommand{\tabref}[1]{Table~\ref{#1}}

\newcommand{\tnote}[1]{} %{\textbf{(T)}\footnote{\textbf{(T)} #1}}

\newcommand{\tdef}[1]{\emph{#1}} %{\textsc{#1}}

\newcommand{\tpre}[1]{}
\newcommand{\tprenote}[1]{}

\providecommand{\url}[1]{#1}
\providecommand{\href}[2]{#2}
\providecommand{\eprint}[1]{\href{http://arXiv.org/abs/#1}{\texttt{arXiv:#1}}}
\newcommand{\Ccal}{\mathcal{C}}
\newcommand{\Dcal}{\mathcal{D}}
\newcommand{\Pcal}{\mathcal{P}}
\newcommand{\Tcal}{\mathcal{T}}
\newcommand{\Vcal}{\mathcal{V}}

\begin{document}

%%%%%%%%%%%%%%%%%%%%%%%%%%%%%%%%%%%%%%%%%%%%%%%
%
% Comment this part out for submission to Scientometrics
%  and ALSO make sure you uncomment the \maketitle command below
%

\renewcommand{\thefootnote}{\fnsymbol{footnote}}

%\title[Was Thebes Necessary?]{Was Thebes Necessary? Contingency in Spatial Modelling}

\begin{center}
{\Large\textbf{Was Thebes Necessary? \\ Contingency in Spatial Modelling\footnote{\texttt{Imperial/TP/16/TSE/2}. \eprint{yymm.nnnnn}. \tsecompldate.  Submitted as part of the Research Topic entitled \href{http://frontiersin.org/Digital_Archaeology/researchtopics/Network_Science_Approaches_for_the_Study_of_Past_Long-Term_Social_Processes_1/4821}{``Network Science Approaches for the Study of Past Long-Term Social Processes''} hosted by Sergi Lozano, Tom Brughmans, Francesca Fulminante and Luce Prignano in Frontiers in Digital Humanities, section Digital Archaeology.} }}
 \\[0.5cm]
 {\large Tim S.\ Evans\,$^{1,2,*}$ and Ray J.\ Rivers\,$^{1,2}$}
 \\[0.5cm]
 $^{1}$\href{https://www.imperial.ac.uk/complexity-science}{Centre for Complexity Science}, Imperial College London, London, SW7 2AZ, U.K.
  \\[0.5cm]
 $^{2}$\href{http://www.imperial.ac.uk/theoretical-physics}{Theoretical Physics}, Physics Dept., Imperial College London, London, SW7 2AZ, U.K.
\end{center}

\providecommand{\title}[1]{}
\providecommand{\author}[1]{}
\providecommand{\institute}[1]{}
\providecommand{\keywords}[1]{\vspace*{12pt} {\centering Keywords: \parbox[t]{0.8\textwidth}{#1}}}
\providecommand{\subclass}[1]{} %{{\centering MSC: \parbox[t]{0.8\textwidth}{#1}}}
\providecommand{\PACS}[1]{} %{{\centering PACS: \parbox[t]{0.8\textwidth}{#1}}}
\providecommand{\date}[1]{}

%
% End of section to comment out for Scientometrics
%
% %%%%%%%%%%%%%%%%%%%%%%%%%%%%%%%%%%%%%%%%%%%%%%%%%%%%%%%%%%%%%%

\begin{abstract}
When data is poor we resort to theory modelling. This is a two-step process. We have first to identify the appropriate type of model for the system under consideration and then to tailor it to the specifics of the case. To understand settlement formation, which is the concern of this paper, this not only involves choosing input parameter values such as site separations but also input functions  which characterises the ease of travel between sites. Although the generic behaviour of the model is understood, the details  are not. Different choices will necessarily lead to different outputs (for identical inputs). We can only proceed if choices that are `close' give outcomes are similar. Where there are local differences it suggests that there was no compelling reason for one outcome rather than the other. If these differences are important for the historic record we may interpret this as sensitivity to contingency. We re-examine the rise of Greek city states as first formulated by Rihll and Wilson in 1979, initially using the same `retail' gravity model. We suggest that, whereas cities like Athens owe their position to a combination of geography and proximity to other sites, the rise of Thebes is the most contingent, whose success reflects social forces outside the grasp of simple network modelling.


%\subclass{91D30} %MSC codes

%\PACS{89.75.-k, 89.75.Hc, 89.65.-s} %PACS codes

\end{abstract}

\renewcommand{\thefootnote}{\arabic{footnote}}
\setcounter{footnote}{0}

% ************************************************************************
\section{Introduction}\tnote{The introduction should be succinct, with no subheadings. Refer to  \href{http://home.frontiersin.org/about/author-guidelines}{Author Guidelines} for further information on how to organize your manuscript in the required sections or their equivalents for your field. For Original Research articles, please note that the Material and Methods section can be placed in any of the following ways: before Results, before Discussion or after Discussion.}

Data is never perfect. However, when it is good, as is often the case in the physical sciences, there are various programmes for tackling data deficiency \citep[e.g.\  ][]{KO01,BKS14}. In social science, and in archaeology in particular, data is often too poor for these approaches to be applicable as they stand. The information we have is incomplete and fragmentary, biased by what has survived, by what has been investigated and what has been made publicly available.  We have to supplement the data that we possess with informed  guesswork of various levels of detail and reliability. In the light of this, in order to make the best use of our limited knowledge or, complementarily, the best use of our ignorance a good starting point is to adopt the fall-back position of answering the question ``All other things being equal, I would expect \ldots to have happened''?

This is where modelling can add to the debate. How to make the best use of our ignorance is an old problem and we shall not attempt to document its history much beyond the observation that economists, who popularised the approach in the 20th century, refer to it \citep[see][chapter 4]{K21} as Laplace's `Principle of Indifference',  although Laplace himself termed it the `Principle of Insufficient Reason' \citep{L25}.  However named, the basic idea (which precedes Laplace) is simple in principle and is exactly how a wise card-player would have played at the gambling tables of the {\it ancien regime}.  List all the `worlds' (in this case, hands of cards) which are compatible with your knowledge/ignorance. Each world is equally likely (or information is being withheld) and the most typical of these is the way in which the system is most likely to have behaved. In the language of Bayesian statistics (Laplace's analysis rather subsumed the ideas of Bayes, but it is Bayes whose name is invoked by statisticians) we have assumed a flat Bayesian prior. Necessarily our conclusions will have a large degree of uncertainty about them and our main aim in this paper is to attempt to show how this uncertainty can be estimated.

This is a basic issue here that sets us, as archaeologists, apart from historians, particularly where data is poor. This is that modelling of the type that we are discussing applies only to generic societal behaviour, behaviour that arises when `history is idling'.  While is is easy to discount the historical staples of major social unrest, famine and disease it is less easy to estimate those aspects of local alliances that lead to one decision rather than another. That is, in the context of the above, we are less interested in an enraged gambler overthrowing the table than in the presence of organised card-sharps. To take the example of this paper that we shall develop at great length later, the formation of Greek city-states in ninth and eight century BC, we shall see that there is a contingency to the presence of Thebes that does not apply to Athens. Indeed, it is hard to get Thebes to appear as a major site. We would see this  as a consequence of social forces (not necessarily dramatic) that modelling of this type, based on broad statistical inference, cannot accommodate. Whereas to a historian the absence of Thebes in its historical position is nonsense, a Bayesian is happy with a local proxy which differs in detail and position from the original, provided something sensible is said about Athens, Corinth and other important sites.

Jaynes \citep{J57,J73} reformulated the Bayes/Laplace approach as the Principle of Maximum Ignorance (or Principle of Epistemic Modesty) and this is the view closest to the methods we will use in this work, in that it can be reformulated as the Principle of Maximum Entropy.  Here we are thinking of entropy in the context of information theory, in which it is simply related to the number of  questions with which we need to interrogate the system to have complete knowledge of it.
Maximum entropy states are, roughly speaking, the ones which have the largest number of ways of creating a result with the specified macroscopic observables (our limited knowledge) and this fits in with Bayes/Laplace.
%One further point: Bayesian/Laplacian notions of `likelihood' are used systematically by astrophysicists studying the coming into existence of the universe and its evolution [e.g. Coles]. Unlike frequentist scientists, who obtain their statistics by repeating their experiments, the universe only happened once. So did history (albeit 13 billion years later) and to try to estimate likelihood in a Bayseian/Laplacian way is a plausible way to proceed.
As always, there are certain caveats that we shall not address and leave to the interested reader \citep{NJ14}.

Our focus here is on models for networks of exchange in the protohistoric past where the primary information contained in the model is spatial. Our worlds are patterns of exchange and a typical use of such models is to see to what extent spatial features alone can account for these.  The entropy maximising that we shall discuss here is only one of several approaches which have evolved in part from urban planning, migration/commuting and economics, for all of which exchange is a key component. We have delineated some of the alternative lines of approach elsewhere \citep{ERK11} and will not pursue those further.

To examine the issues with quantifying levels of expectation with their concomitant uncertainty and ambiguity, we will build on a classic study using entropic methods by Rihll and Wilson (\citeyear{RW87,RW91}). They examined how spatial models, based on models for siting retail outlets \citep{H64,LH65,HW78},\tnote{See pp.11 of \citet{RW87} for further citations.} can be used to study the rise of city states in central mainland Greece around the eighth century BC. Specifically, the question is to understand to what extent spatial features can explain why certain settlements out of the many in the region came to dominate it.
The success of the model led to its adoption for understanding city state formation in other times and other places; 2nd millennium BC Crete \citep{BW13}, Middle Bronze Age and Iron Age NE Syria \citep{DFWPAR14}, early 2nd millennium BC Central Anatolia \citep{PA15} and late 1st Millenium Latenian urbanisation \citep{F16a}.

Our aim is, using the data of Rihll and Wilson as the example, to show how to improve the treatment of uncertainty in such spatial modelling. This lies behind the title of this paper. We shall show that some sites, like Athens, are likely to be significant despite our poor knowledge whereas some, like Thebes, are much more contingent on information that we don't possess. Unlike Athens, its historical importance is likely a consequence of social forces that models like ours cannot accommodate. We had suggested this previously \citep{RE14} and this current paper provides an extensive development and analysis of this earlier proposition. However our focus in this article is less on the archaeology of late Geometric/Early Archaic Greece and more on the nature of uncertainty in such approaches. As such it is equally applicable to the other case studies and some preliminary work on Latenian urbanisation is under way.

As we have said, these studies of urbanisation are framed in the language of exchange networks. In the first instance `exchange' means the physical transportation of things and people, with secondary meanings in terms of power and culture that these things and people embody, the details of which we know very imperfectly. Uncertainty comes in many forms but, roughly, it can be divided into uncertainty in `physical' parameters and uncertainty in `calibration' \citep{KO01}.  For the former a major source of uncertainty lies in our imperfect knowledge of site locations. In this regard, we take the sites of the Rihll and Wilson model as given. Even then, further uncertainty lies in the nature of the paths taken (we are thinking here of land-based and riverine travel) e.g. to what extent do existing roads and waterways determine long-distance exchange? We consider the effect of adopting different definitions. A third problem arises in estimating the effort/cost/time involved in the transportation along these routes. These, plus the ambiguity in the site carrying capacities or resource bases, constitute the major uncertainties in the physical parameters of the model. As for calibration, there is the question of how to characterise the ease of exchange for different `costs' or distances. This is encoded in the so-called `deterrence' function, a calibration function, with its own calibration parameters, which imposes cost penalties on long-distance single transactions. We consider the effect of different choices of this deterrence function. In fact, what began our analysis of this data set \citep{RE14}  was our inability to replicate the results of Rihll and Wilson when using a different choice of deterrence function from them.

Superficially, our conclusions seem very plausible just on geometric grounds from the distribution of sites. This leads us to question whether we needed the apparatus of entropy and Bayesian analysis. Could we have reached good enough conclusions with a ruler and compass? To test such a null approach we compare the results of this model to the results obtained using more traditional data clustering methods, which rely on geography alone. We find that the answer is no. As part of this programme we produce an open source list of site locations and distances used which allows future researchers to test their ideas on this data sets.

There is the caveat that even seemingly epistemic approaches like this can have a manifestly ontic realisation in terms of sites as actors in their evolution. For example, Wilson rephrases the model in terms of dynamical Lokta-Volterra (predator-prey) equations \citep{W08}, and Altaweel has re-examined Syrian city-state formation from an agent-based approach \citep{A15}.  Such approaches provide useful complementary perspectives.

% ************************************************************************
\section{Data}

% -------------------------------------------------------------------------------
\subsection{Sites}\label{sssitedata}

\begin{figure}[h!]
 \begin{center}
  \includegraphics[width=16cm]{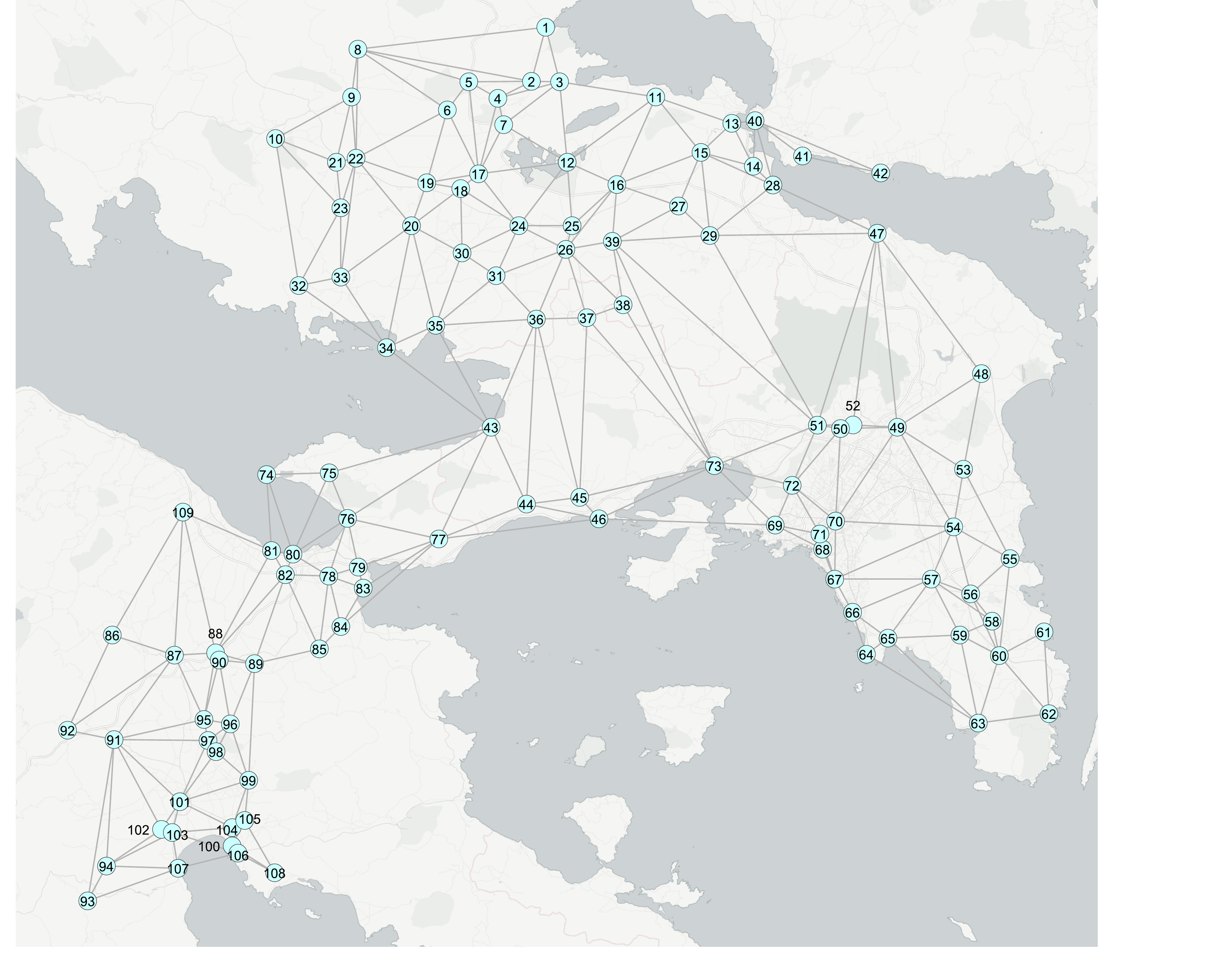}
 \end{center}
 \caption{The approximate locations of the 109 sites used as the starting point for this study, derived from Fig.~1 of \cite{RW87}. The index of site numbers and the precise locations used in this study are given in \tabref{tRW109index} in the Supplementary Material. Note that site 64, Vouliagmeni, was not labelled in the original figure, see \citet{E16a} for more details.  The `normal' distance set are the direct straight line distances between these sites and the full table of distances are given in \citet{E16a}. The edges are a subset of those given by Delaunay Triangulation and are used to calculate a second set of distances (denoted DTMedit). Each edge is linked to the straight distance between the end points.  The distance between other pairs of points is found by taking the length of the shortest path along the edges where the path distance is the sum of the distances associated with the edges traversed in the path.}\label{fRW109numbers}
\end{figure}

The primary data we use is that of Rihll and Wilson (\citeyear{RW91}, pp.68-71); namely the choice and location of sites.  The sites chosen constitute 109 population centres from the late Geometric period, the ninth and eight century BC,\tnote{Geometric is used on p68 of \citet{RW91} while 8th and 9th c.\ BC is used on page 12 of \citet{RW87}.}in a region roughly 130km across, as specified in Fig.~1 of \citet{RW87} and shown here in \figref{fRW109numbers}. Some of the hypotheses behind the choices are outlined in \citet{RW87}, e.g. using the existence of a temple or the size of a site as a marker of importance.
%\tnote{A detailed explanation of why these sites were included and why unspecified other sites were excluded is not provided in \citet{RW87} or \citet{RW91}.}
Relying on such expert judgement is inevitable, but this approach immediately provides a largely unquantifiable source of uncertainty. One response would be to try different `reasonable' choices of sites to see what effect this has on conclusions.  However we will not try this here and we take this set of 109 sites in \figref{fRW109numbers} as given.

Some models can accommodate some sense of the size of the site. However Rihll and Wilson judge that this information is very uncertain in this period and it is more reasonable to assume all sites were roughly equal in size and importance at the start of the period \citep[see][pp.69-70]{RW91}.  This reflects a entropy viewpoint that this is the least biased assumption to make when there is no other information available.

There is one obvious feature about the spatial arrangement of these 109 sites and that is there is a natural and obvious division into three regions: Boeotia, Attica and the Isthmus-Argolid region. We will comment further on the properties of the spatial arrangement of these 109 sites below, see section \ref{ssregions}.

% -------------------------------------------------------------------------------
\subsection{Distances}\label{ssdistancedata}

The only other explicit information used here, and in the original papers, is the distance between every pair of sites \citep[see][pp.63-65]{RW91}. There are many different ways to assess the effective separation of two sites: effort, time of travel, financial costs, distance of the actual route followed,  etc.  For our context, only the last can be measured with any accuracy using modern GIS methods, but lack of knowledge about the details of exchange and the way landscapes were actually used still leaves unquantifiable uncertainties.  Again the best response is to make few assumptions so in the first instance, like Rihll and Wilson, we use a straight-line distance between sites.  In fact the choice of significant sites may well contain some implicit information about the landscape, e.g. an area of difficult terrain is likely to have fewer settlements. We will try to judge the effect of the uncertainty of distance as part of our study, one aspect where we extend the original work of \citep{RE14}.

One complication when comparing our results to the original studies is that the distances used in the latter are not provided.  To obtain a comparable set of distances we have digitised Fig.~1 of \citet{RW87}; our coordinates for the sites are given in the Supplementary Material in \tabref{tRW109index}. The distance between sites is then the length of a straight line between these sites in these coordinates. This comprises what we will call our `normal' distance data set. This data is given in \citet{E16a} and the process is discussed there in more detail.
The scale is that roughly 6 of our units are equivalent to 1km. If we look at the nearest neighbour to each site, we find that the closest pair are 8 units apart (Nauplia - Pronaia) and the furthest distance between nearest neighbours is 75 units (from Sikyon to Akraia). These minimum distances have a first quartile of 20.0, a median of 28 (e.g.\ Kopai-Olmous), and a third quartile of 39.0. The mean is 30.9 units\tnote{Use SpatialAnalysis.distanceAnalysis.py} implying a short distance scale of around 5km.

Perhaps the most obvious problem with this data is that the shortest path between many pairs of points includes paths which are over the sea. Travelling on water in the ancient world was, where it was available, the most effective method of long distance communication, both in terms of speed and in terms of bulk. However land and water borne transport have different advantages and disadvantages meaning that we cannot simply mix these two modes of travel.  This issue largely effects long range links across our region.  We shall see later that the bulk of the results seem to be sensitive to smaller scales, say 30km or less (the maximum walking distance in one day).  We might reasonably hope that sea travel is likely to have little impact on our results and have used this to produce a second set of separations based purely on routes which avoid the open sea.

From the above we see sites are so close that a determined individual could visit several sites in a day, most likely  making use of an existing system of tracks/roads, which we assume occur between nearest neighbours. Thus, if going from site A to site C takes us near an intermediate site B, the traveler is more likely to go from A to C via B than strike off a direct route to save a short distance. Given the uncertainty in assigning useful distances between these ancient sites, we judge that any differences in the distances used here as compared to the original studies should be no more significant than many of the other sources of uncertainty. To test this  we applied Delaunay Triangulation to the same 109 sites so as to produce a set of non-interscting edges between `nearest neighbour' sites. Again, for sites around the edge of the region, this produces several long distance links which cut across the sea.  We choose which links to remove using our judgement, ending with a set of largely land based links.  The distance for every remaining link was exactly as before, with a typical separation between nearest neighbours of around 5 to 10 km.  We then found the distance between each pair of sites by assuming that the path between sites will always be along the links between nearest neighbours, using the path with the shortest total path length.  This Delaunay Triangulated derived distance set (labelled `DTMedit' in figures) is shown in \figref{fRW109numbers}.

In practice this means that sites which are not nearest neighbours are actually slightly further apart than in our normal data set. Again this process is not precise, but the differences between the set of Delaunay Triangulation derived distances and our normal distance set (direct paths) can serve as a proxy for the sort of errors we might expect in our estimations of effective distance between sites.
\tnote{Other ways of probing the effects of this source of uncertainty are easily imagined e.g. adding some random noise to the original distances, but that is left for future work. Add a picture to illustrate (a) nn, (b) removed link, (c) path between non-nearest neighbours.}
The differences between our two different sets of distances and the Rihll and Wilson direct path distances will serve as a test of the robustness of results against uncertainty in travel.

\begin{figure}[h!]
 \begin{center}
  %Figure showing HAC for RW109b and RW109bDTMedit
  \includegraphics[width=16cm]{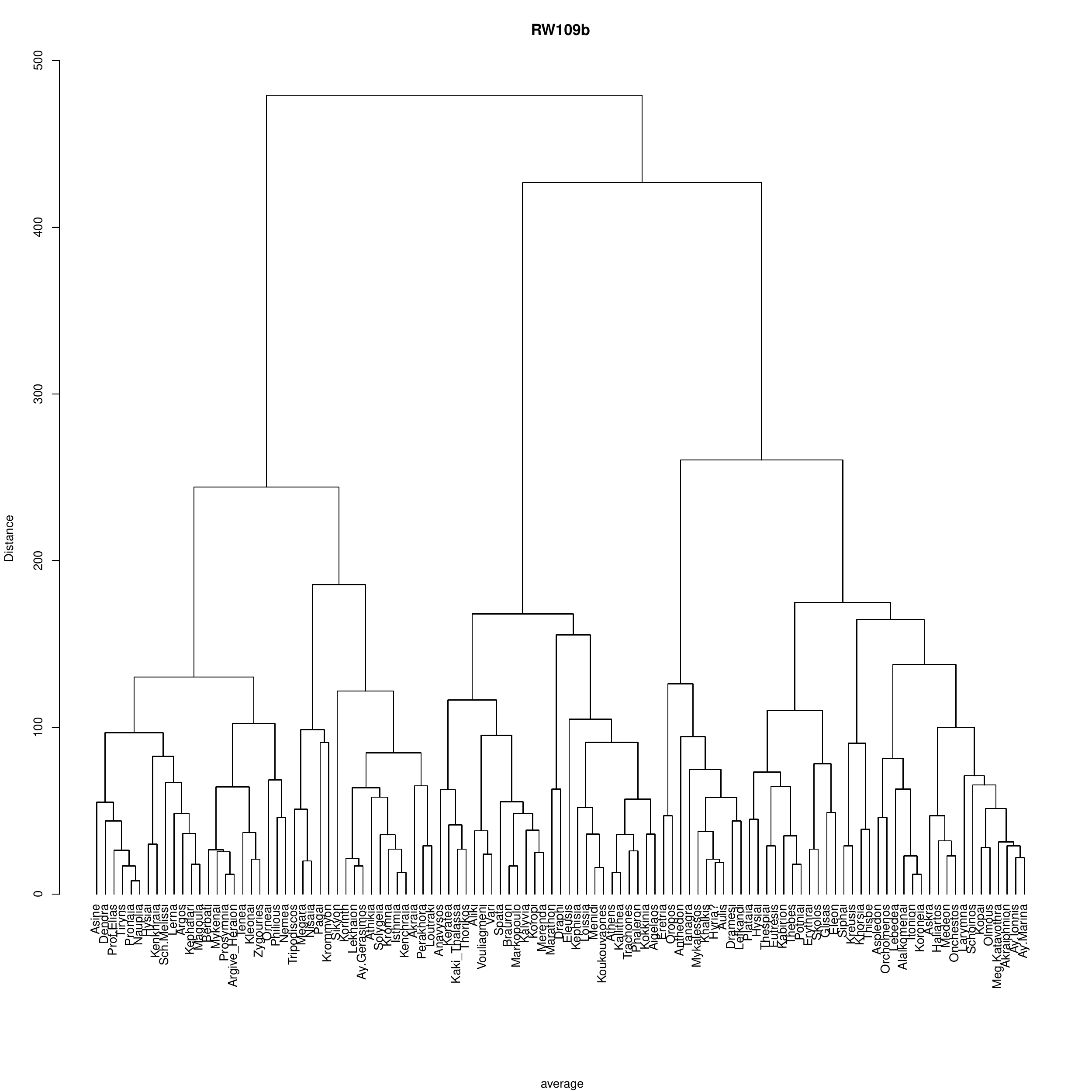}% This is a *.jpg file
 \end{center}
 \caption{Hierarchical Agglomerative Clustering for the normal distance set using the average criterion for agglomeration.}\label{fhacnormal}
\end{figure}

% ------------------------------------------------------------------
\subsection{Regional structure of the data}\label{ssregions}

It is worth looking at our distance sets for obvious features.  Using Hierarchical Agglomerative Clustering \citep[for example see][]{MRS08}, a standard method to cluster the data,  we find that there are only really three main scales in either distance data set.  The typical intersite separation marks the onset of clustering, the overall size of the data set (around 130km) marks the end.  In between there is only one characteristic scale in the dendrogram of \figref{fhacnormal} (see also \figref{fhacDTMedit} in the Supplementary Material).  This is a region of three clusters, which reflect the regions delineated by clear gaps in the map of sites, as seen in \figref{fRW109numbers}.  They correspond to Boeotia in the north (containing Thebes), Attica in the east (containing Athens), and the Isthmus/Argive region in the south west (containing Corinth and Argos). While the two distance data sets produce different dendrograms in detail, the large scale picture is broadly similar for both.
\tnote{TODO: Could plot size size vs.\ site rank (smallest distance first) to show this.}
%\tnote{ToDo?
%Sparsification for RW model Thresholding Edge present if flow bigger than some value F0. Not tried this   but may well work as (a) typical RW site has one strong edge out, (b) a few RW sites are star like with several equally weighted out edges.}

% *********************************************************************************
\section{The Rihll and Wilson model}%\tnote{Do we want to find a more generic name}

The model used by Rihll and Wilson \citeyearpar{RW87,RW91} and by later authors \citep{WD10,BW13,RE14} to describe urbanisation and state formation in historical contexts was originally devised to study the emergence of dominant retail centres \citep{H64,LH65,HW78}.
The model differs from several other well known spatial models, such as Gravity Models, in that it is a `Zone of Control' model in the terminology of \citet{ERK11}. That is the model produces a structure of dominant sites, here called `terminal sites', surrounded by a cluster of smaller nearby satellite sites, see Fig.s \ref{fRWGMPEXPb1050d85} and  \ref{fRWGMPEXPb1050d90}.

In all that follows we consider a network of $N$ sites labelled $i= 1,2,... ,N$. As mentioned above in section~\ref{sssitedata}, we take the sites to be equal, differing only in terms of their locations relative to one another. The output of the model will be a set of `flows' $F_{ij}$ each of which describes a flow from site $i$ to a distinct site $j$, separated by an appropriately defined distance $d_{ij}$.  In many spatial models this flow $F_{ij}$ represents the transmission of goods, people, influence, ideas etc.\ from site $i$ to site $j$. Here however we will use it to represent the ``pulling power or attractiveness'' \citep[pp.8]{RW87} of site $i$ to site $j$. This is, of course, very abstract and unquantifiable but in most ancient contexts it is nearly always impossible to quantify exchanges of any type between sites, even where they are physical goods, so this is not a particular weakness of this study.

% ---------------------------------------------------------------
\subsection{Construction of the Rihll and Wilson model}

In the Supplementary Material we outline the derivation of the model though the maximum entropy approach of \citet{W67}.  The key assumption is that if all other things are equal, then every possible exchange counted by the flows is equally likely.  Of course in reality there are strong constraints. We impose these here in three steps producing first the simple gravity model, then the output constrained gravity model and finally, the Rihll and Wilson model. We shall avoid algebra as much as possible here and refer the reader to the Supplementary Material, where it is given in greater detail.

% .................................................................
\subsubsection{The Simple Gravity Model}

Exchange between sites $i$ and $j$ requires some effort and we imagine that this can be roughly quantified as a `cost' $c_{ij}$.  For instance the effort to maintain a link between two sites will generally increase with separation $d_{ij}$ so we might choose to capture this with the simple choice that our costs are equal to the distances, $c_{ij}=d_{ij}$. Our single constraint is that the total cost/effort that can be expended on exchange is capped. This is natural given that any society has limited resources. Making the best use of this knowledge (maximising the entropy of the exchange flows subject to this constraint) gives the most likely distribution as
\beq
 F_{ij}= A \, f(d_{ij}/D) \, , \qquad
 i \neq j \, .
\label{FRW0}
\eeq
Here the constant $A$ is determined by the total cost we allow but it can be fixed more easily by specifying the total amount of exchange, $F_\mathrm{total} =\sum_{i,j} F_{ij}$.
The function $f(x)$ is known as the \tdef{deterrence function} and it expresses the effect of the costs incurred on a trip from $i$ to $j$ in terms of the distance.  The likelihood of a single exchange occurring over distance $d_{ij}$ is proportional to $f(d_{ij}/D)$. The new parameter $D$ is a calibration scale for the effect on distances on travel. Often $f(x)$ is chosen such that $f(0) = 1$.
The appropriate form for the deterrence function is rarely known so another source of uncertainty comes from its choice.  Note that we have swapped our lack of knowledge about the costs $c_{ij}$ for the lack of knowledge about the function $f$.  The gain is that we have reexpressed the aspect in terms of a function of distance.  Distances are more accessible quantities than costs and we have some knowledge, or at least intuition, about the effect of distance on exchange. For instance we imagine that the deterrence function should always decrease as distance gets larger and indeed all our examples have this simple feature.
Part of our study is to see how resilient the results from our modelling are to different choices.

A common form for the deterrence function corresponds to the simplest choice that costs are proportional to distances.  This leads to the simple exponential form (labelled EXP in figures)
\beq
 f(d_{ij}/D) = \exp(-d_{ij}/D ) \, .
 \label{detfuncexp}
\eeq
This is also the form used in \citeauthor{RW87} \citeyearpar{RW87,RW91}.

\begin{figure}[h!]
 \begin{center}
  %Figure showing potentials
  \includegraphics[width=10cm]{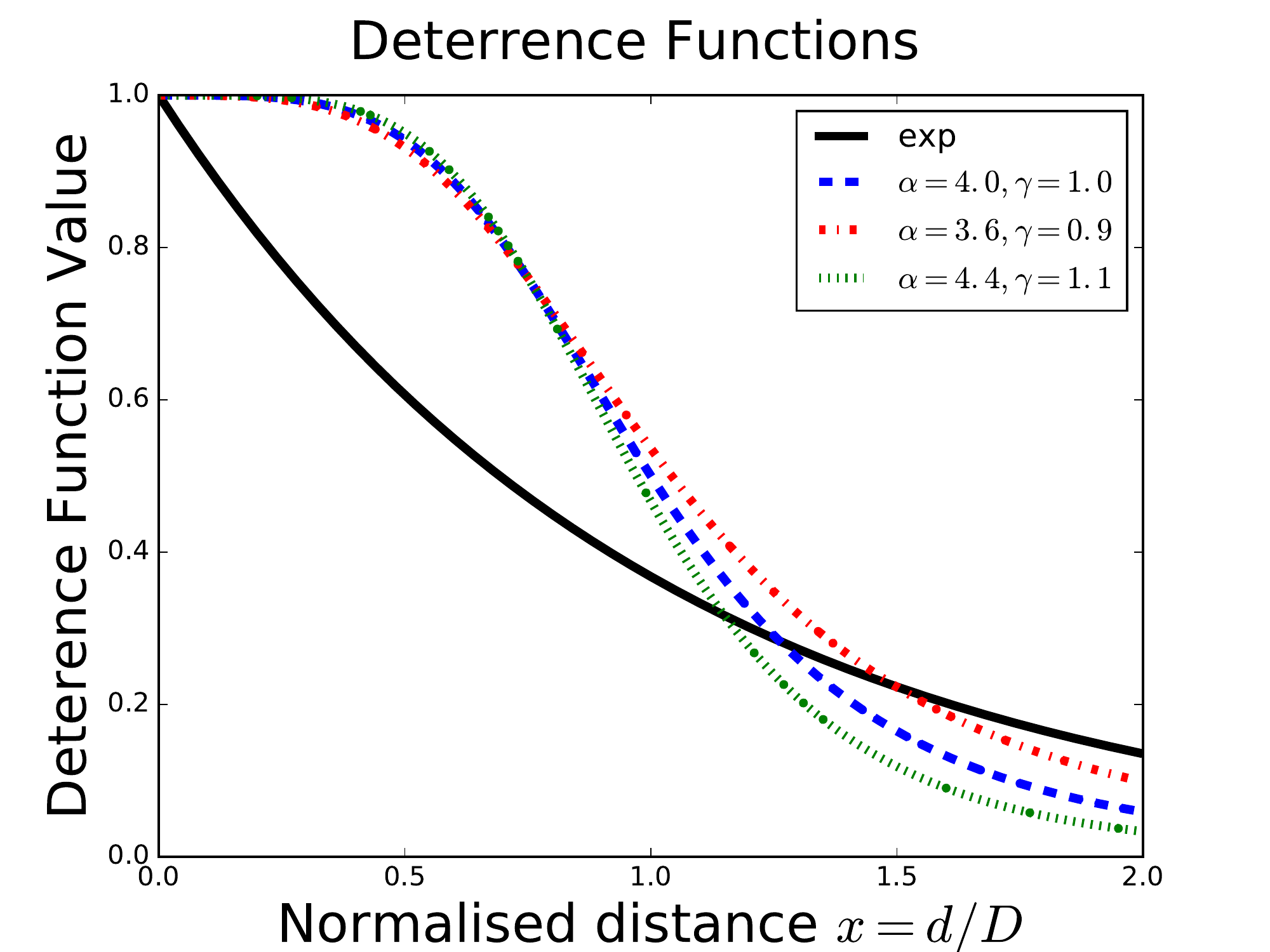}
 \end{center}
 \caption{Figure showing exponential deterrence function (EXP) of \tref{detfuncexp} and the `ariadne' deterrence function of \tref{detfuncaep} in terms of $x=d/D$, the distance $d$ scaled by the distance parameter $D$. The ariadne deterrence functions \tref{detfuncaep} shown for three of the nine parameter value combinations $(\alpha,\gamma)$ used in this work.}\label{fpotl}
\end{figure}

In general the deterrence function can depend on additional calibration parameters that alter its shape, as we see in the second form shown in \figref{fpotl}.  This `ariadne' deterrence function (labelled AEP in figures and used in  \citet{EKR06}, \citeauthor{KER08} \citeyearpar{KER08,KER11}, \citet{ERK11}), might seem more appropriate to the case in hand where the proximity of sites suggests low penalties for short distance exchange.\tnote{Citation/explanation here???}
\tnote{(this is also appropriate to water-bourne exchange with loading and unloading costs).  Not sure this is relevant here.} It takes what we term the \tdef{ariadne} form
\beq
 f(d_{ij}/D) = \left( 1+ (d_{ij}/D)^\alpha\right)^{-\gamma} \, .
 \label{detfuncaep}
\eeq
As well as the calibration distance scale $D$, this form has two further parameters, $\alpha$ and $\gamma$, which alter the shape of the function. We will work with $\alpha=3.6$, $4.0$ and $4.4$ and while we will use $\gamma$ values of $0.9$, $1.0$ and $1.1$.

% .................................................................
\subsubsection{The Output Constrained Gravity Model}

The Simple Gravity model as given above can also be derived from a manifestly ontic, agent based approach \citep{J72}. However, the Simple Gravity model is deficient in many ways.

One clear limitation is that there is no particular limit on the amount of exchange emerging from each site in the Simple Gravity model, only the total amount of exchange is limited. By way of comparison we note that some of the earliest and most successful examples of spatial modelling in archaeology have employed Proximal Point Analysis \citep[for example][]{T77,B00}. Proximal Point Analysis posits that each site only has interactions/exchanges with a fixed number of their closest sites, i.e.\ outflows are constrained. This has parallels in the idea that individuals only have the capacity to really interact with a limited number of other individuals \citep{D92}.

Adding to the Simple Gravity model the constraint that individual site outflows are fixed to be a given value $O_i$ produces the Output Constrained Gravity model where the most likely exchange flows are found to be
 \beq
 F_{ij}=A_i O_i f(d_{ij}/D)    \, , \qquad
 A_i^{-1} = \sum_j f(d_{ij}/D) \, , \qquad
 i \neq j \, .
\label{FRW}
\eeq
The normalisation factors $A_i$ ensure that the output from each site is indeed $O_i$, that is $O_i = \sum_j F_{ij}$.

% .................................................................
\subsubsection{The Rihll and Wilson model}

Unless we extend these models to accommodate strongly unequal site sizes initially, they have no mechanism for generating a handful of dominant sites as outputs, as identified in the archaeological record.  The presence of dominant sites requires the inclusion of non-linear behaviour in the model constraints.
Drawing on earlier work modelling retail outlets \citep{H64,LH65,HW78}, Rihll and Wilson do this by constraining the entropy of the site inflows $I_j = \sum_i F_{ij}$ though the motivation for this is best provided posthoc from the form of the final model. In the absence of this input entropy constraint, inflows over similar distance scales tend not to have wide variation as seen in either of the two previous models. The effect of this input entropy constraint is to give non-linear feedback so that sites which develop an advantage build on that advantage to become even more important, suggestive of the synoikism and urbanisation that is expected to lie behind the appearance of regionally dominant sites. It is the inflow $I_j$ {\it outputs} determined by the model that are used by Rihll and Wilson to assign an importance or `attractiveness' to a site\footnote{In a second self-consistent version, Rihll and Wilson set the outputs equal to the inputs in a self-consistent manner $O_i=I_i$ so the number of arrivals is the number of departures. The model parameters now include an initial value for $I_i$ for each site.  We will not consider this self-consistent variant further.}.

The result (see supplementary material section \ref{sRWtech}) is that the flow $F_{ij}$ from site $i$ to site $j$ ($i \neq j$) now takes the form \footnote{Note that it is standard to ignore the internal flows, the entries $F_{ii}$. Equivalently we choose $f$ so that $f(d_{ii}/D)=f(0)=0$.}
\beq
 F_{ij}=A_i O_i I_j^\beta f(d_{ij}/D)    \, , \qquad
 A_i^{-1} = \sum_k I_k^\beta f(d_{ij}/D) \, , \qquad
 i \neq j \, .
\label{FRW2}
\eeq
As before, the normalisation factors $A_i$ ensure that the output from each site is indeed $O_i$. We stress that the total flows \emph{into} each site $j$, the $I_j$, are not parameters of the theory but are all specified by the solution.
The cost of each exchange is encoded in the deterrence function, $f(d_{ij}/D)$, as before.
Again the model demands that the outputs are fixed.
Given our limited information, we are unable to specify the outflows on a site by site basis and we follow our usual response to a lack of information and take the outflows to be equal in this case. Rihll and Wilson also set the output from each site to be equal  as they suggest it is most reasonable to assume all sites were roughly equal in size and importance at the start of the period \citep[see][pp.69-70]{RW91}.

\begin{figure}[h!]
 \begin{center}
 % Figure produced from terminalAnalysis.py using data in terminalnumberb1050.dat derived from RW109bEither_RWGM_PEXPAEP_mastats.xlsx
  \includegraphics[width=16cm]{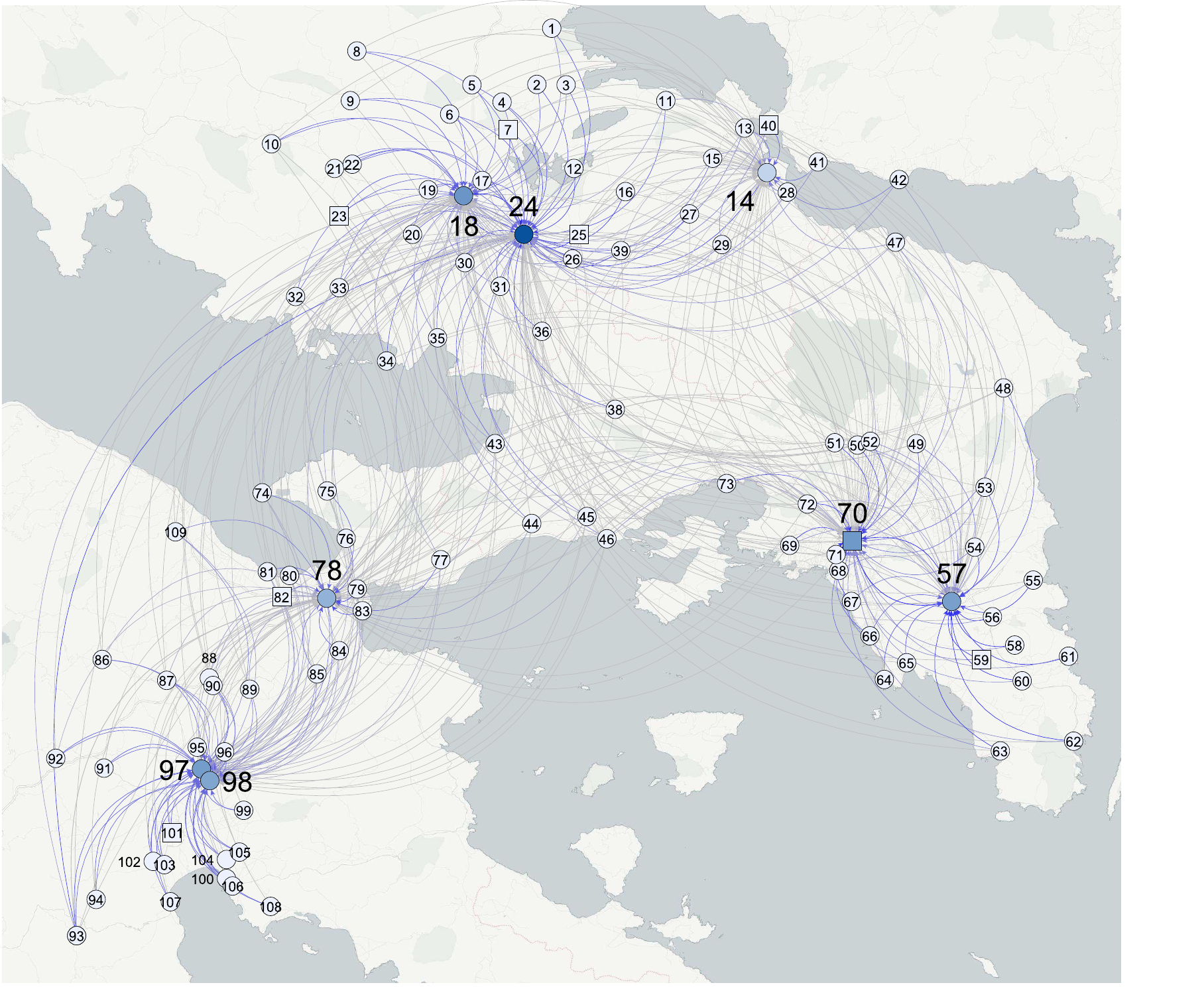}
 \end{center}
 \caption{Figure showing a solution for the Rihll and Wilson model \tref{FRW} with an exponential deterrent function, $D=85$, $\beta=1.05$ using normal (direct) distances. The darker the colour of an edge the large the flow along that edge. The square sites are the terminal sites shown in fig.6 of \citet{RW91}.}\label{fRWGMPEXPb1050d85}
\end{figure}

\begin{figure}[h!]
 \begin{center}
 % Figure produced from terminalAnalysis.py using data in terminalnumberb1050.dat derived from RW109bEither_RWGM_PEXPAEP_mastats.xlsx
  \includegraphics[width=16cm]{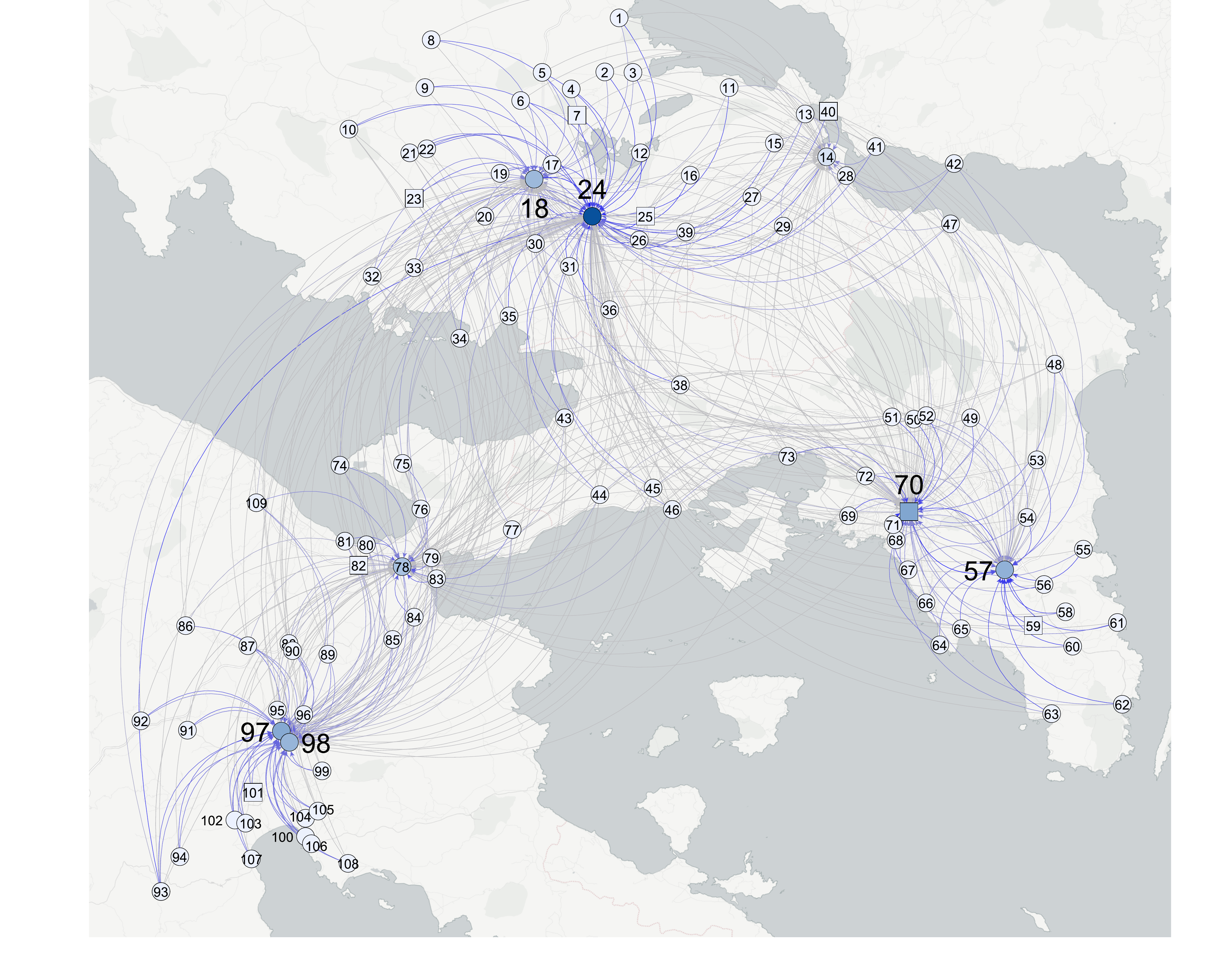}
 \end{center}
 \caption{Figure showing a solution for the Rihll and Wilson model \tref{FRW} with an exponential deterrent function, $D=90$, $\beta=1.05$ using normal (direct) distances. The darker the colour of an edge the large the flow along that edge.}\label{fRWGMPEXPb1050d90}
\end{figure}

The feature which distinguishes this model from other gravity models is the factor of $I_j^\beta$ in \tref{FRW2}.  The $I_j^\beta$ leads to solutions where for most sites $I_j$ is zero or close to zero so that all the available flow is input to a few sites, the ``terminal sites'',  which have non-trivial input flows.  These terminal sites send their output flows to a variety of other terminal sites. Roughly speaking if the deterrence function becomes negligible for sites separated by distance scale $D$ or more, then a site $T$ with a large input flow (or attractiveness) $I_T$ will suppress the attractiveness of sites within a radius of $D$ or so through the normalisation factor $A_i$. Basically a first guess is that in this model \tref{FRW2} the system will split up into patches of radius $D$, each with one dominant site.  Indeed this is the typical pattern of solutions to the Rihll and Wilson model; a series of stars where all the flow leaving most sites is directed to just one site in their neighbourhood, as can be seen in Fig.s \ref{fRWGMPEXPb1050d85} and  \ref{fRWGMPEXPb1050d90}.

\clearpage

% ----------------------------------------------------------------------
\subsection{Terminal sites}\label{ssterminalsites}

The difference from gravity models with input and output constraints \citep[for instance see][]{ERK11}  is that outflows $O_i$ are still {\it input} parameters of the theory but now the inflows $I_j$ are {\it outputs} determined by the model. These are used by Rihll and Wilson to assign an importance or `attractiveness' to a site.

To identify the most important sites, Rihll and Wilson use a particular implementation of a scheme of \citet{ND61} and we will follow suit. We define a \tdef{terminal site} to be a site where the total flow into that site is bigger than the largest flow out of that terminal site along any one edge.  That is a terminal site $T$ satisfies
\beq
 I_{T} = \sum_i F_{iT}  > F_{Tj} \; \; \; \forall \; j \neq T \, .
\label{terminaldef}
\eeq
Rihll and Wilson then compare the terminal sites found from the results of the model with the archaeological record. In fact the nature of the output of the model, one dominated by obvious star like formations which naturally define `zones of control', means that we can pick these terminal sites out by eye from visualisations of the complete flow matrix in most cases, as in Fig.s \ref{fRWGMPEXPb1050d85} and  \ref{fRWGMPEXPb1050d90}.

% ************************************************************************
\section{Results}

%\tnote{Uncertainty should be reflected in results.  Should see ``errors'' - variability measures
 % --- Vary initial conditions ?, Stochastic dynamics ? (but ariadne/ERG), Vary parameters ?, Different Models ?.
% $ Illustration with Rihll and Wilson data (1987, 1991).}

% -------------------------------------------------------------------------------
\subsection{Assessing the results}\label{ssassessresults}

The basic idea in the original studies is to find the sites which emerge from the initial equal sized settlements to dominate a region, based purely on the relative geographical positions.

In this context there are certainly four cities which dominate the history of subsequent periods in this region: Thebes, Athens, Corinth and Argos\footnote{To this list Rihll and Wilson add a fifth site, Khalkis \citep[see][pp.71]{RW91}, which is at the narrowest point between the mainland and the large island of Euboea. Khalkis was significant enough to found several other Greek cities.}. We will use these four well known sites as our key measure of the effectiveness of our various modelling attempts. Of course other sites play important roles; Rihll and Wilson provide commentaries on several other sites, both important and less so (throughout \citeauthor{RW87} \citeyear{RW87} and \citeyear{RW91} but see pp.71 of the latter in particular). Undoubtedly scholarship about the settlements in this region at this time will have moved on but our focus is on the methodology and the role of uncertainty in modelling so we are content to remain with a data set which captures the broad picture. It also enables us to build directly on the specific results present in the original papers.

Rihll and Wilson looked at several parameter values in their 1987 and 1991 papers for the exponential deterrence function of (\ref{detfuncexp}).  Outcomes are determined by two parameters; the distance scale $D$ of the deterrence function and the `attractiveness' exponent $\beta$. Their results for these four significant sites, as derived from their figures, are shown in \tabref{trwterminals}. Thebes and Athens were always identified as terminal sites. Corinth is identified in four of these figures but sometimes the terminal site was not Corinth itself but a close neighbour.  There was also always a terminal at Argos or a close neighbour.

\begin{table}
\begin{center}
\begin{tabular}{c|c|c|c||c|c|c|c|c}
  Paper & Figure & $\beta$ & $D_\mathrm{RW}$ & Number    & Thebes & Athens & Corinth & Argos  \\
        & Number &         &                 & Terminals &        &        &         &        \\
  \hline\hline
  1987 &  2 & 1.010 & 6.667 &  7 & Y & Y &   Y  & (98) \\ \hline
  1987 &  4 & 1.025 & 5.714 &  8 & Y & Y &   Y  &   Y  \\ \hline
  1987 &  9 & 1.025 & 6.667 &  8 & Y & Y &   Y  & (98) \\ \hline
  1987 &  7 & 1.010 & 6.667 &  8 & Y & Y & (78) & (98) \\ \hline
  1991 &  6 & 1.050 & 5.714 &  8 & Y & Y &   Y  &   Y  \\ \hline
  1991 &  5 & 1.005 & 5.714 & 10 & Y & Y & (78) & (98) \\ \hline
  \begin{tabular}{@{}c@{}}1987 \\ 1991 \end{tabular}
       &
  \begin{tabular}{@{}c@{}} 11 \\ 4 \end{tabular}
            & 1.025 & 4.000 & 13 & Y & Y & (78) & (98)
  %\\ 1991 &  4 & 1.025 & 4.000 & 13 & Y & Y & (78) & (98) \\
\end{tabular}
\end{center}
\caption{The list of results for the seven different parameter values shown in the two Rihll and Wilson papers from 1987 and 1991. For each of the four main sites considered, a `Y' indicates that site was a terminal site in the corresponding figure.  A number indicates the index of the terminal site closest to the given city when the latter is not a terminal site in that figure. Site 78 is Kromna which is very close to Corinth (distance 36 in our units) while site 98, the Argive Heraion, is close to Argos (a distance of 37 in our units, roughly 8km in reality). Athens and Thebes are always identified correctly. The distance scale $D_\mathrm{RW}$ is terms of the unspecified units used in the original papers.  This $D_\mathrm{RW}$ is simply the inverse of $\gamma_{RW}$, the scaling factor of distance in the exponential form of the potential \tref{detfuncexp} as it is this $\gamma_{RW}$ value which is the value quoted in the original papers.}\label{trwterminals}
\end{table}

\begin{figure}[h!]
 \begin{center}
 % produced by terminalAnalysis.py
  \includegraphics[width=12cm]{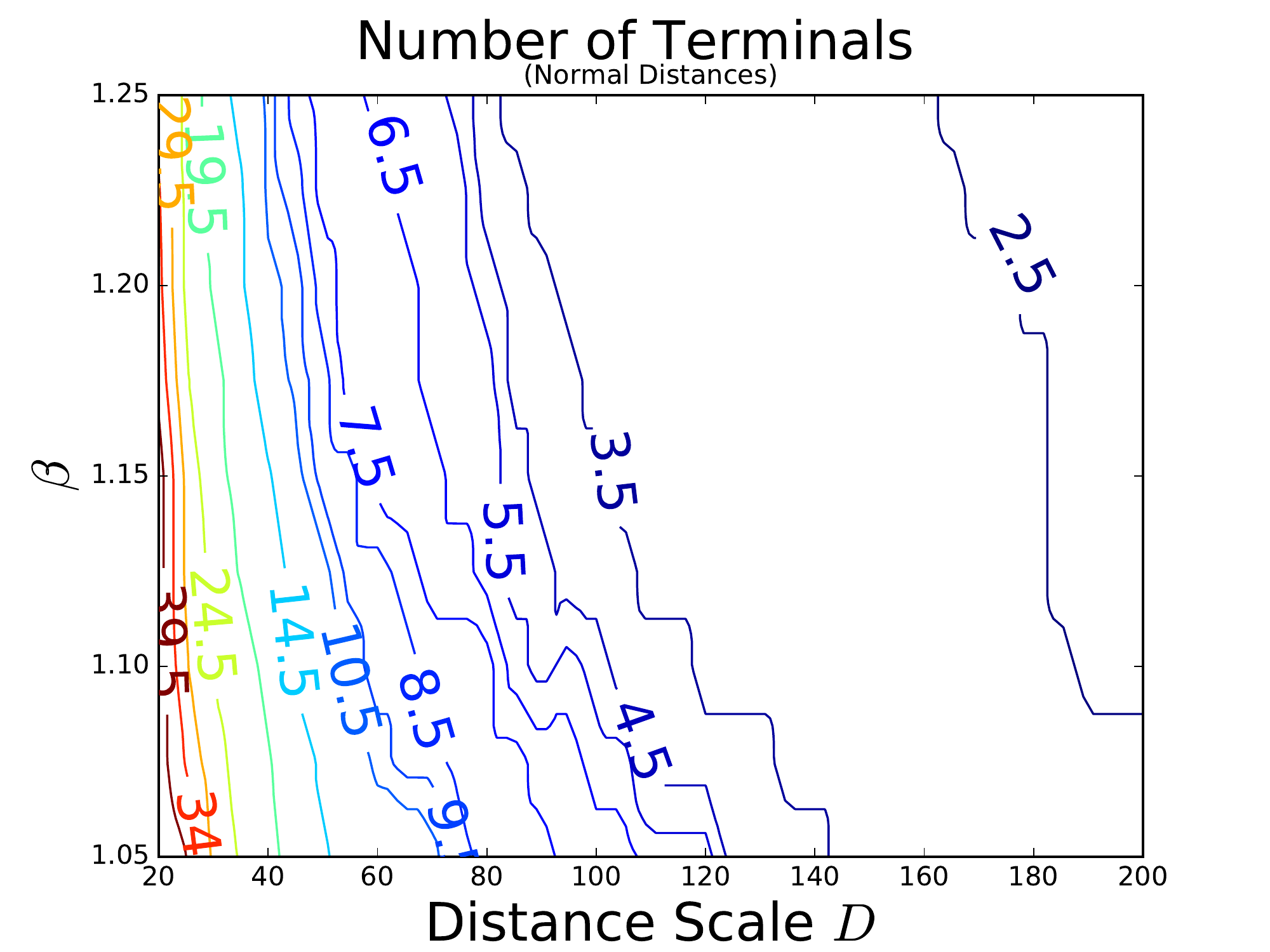}
 \end{center}
 \caption{Figure showing the number of terminals as the distance scale $D$ and the non-linearity parameter $\beta$ are varied. The contours are placed at half integer values as the actual terminal numbers are integers. For exponential potential \tref{detfuncexp} and normal distance set.}\label{fPEXPbvsDTcontour}
\end{figure}

In terms of robustness, Rihll and Wilson look at a range of parameter values, producing between eight and thirteen terminal in the figures 4 to 6 of \citet{RW91}. The key results are given in Table~1.
We repeat this in the first stage of our study by looking at how sensitive our results are to changes in the calibration parameters $D$ and $\beta$ for a given deterrence function. In \figref{fPEXPbvsDTcontour} we show the sensitivity of the terminal number to the distance $D$ and $\beta$ parameter values, for the standard exponential deterrence function and for the normal (direct) distance set.  This is broadly similar to Fig.~6 of \citet{RW87}. Essentially the only solutions which are relatively stable are those with three terminals.  This is clear from the dendrograms of figures \ref{fhacnormal} (see also  \ref{fhacDTMedit} in the Supplementary Material) where as the distance scale is raised, clusters start to form at around the minimum separation scale and then grow without any characteristics scale until they reach three clusters.  There is then a long range of the distance scale where the dendrograms are stable with three clusters.  Normally when clustering data, such a large region of stability is taken to indicate that this is a `good' clustering and the user is confident that these clusters represent a significant feature in the data.  Here this stability with three clusters reflects the only clear spatial feature of our sites, the geographical gap which separates them into the three regions of Boeotia, Attica and Isthmus/Argolid.

% ----------------------------------------
\subsection{Fixed $\beta$}

\begin{figure}[h!]
 \begin{center}
 % produced by terminalAnalysis.py
  \includegraphics[width=12cm]{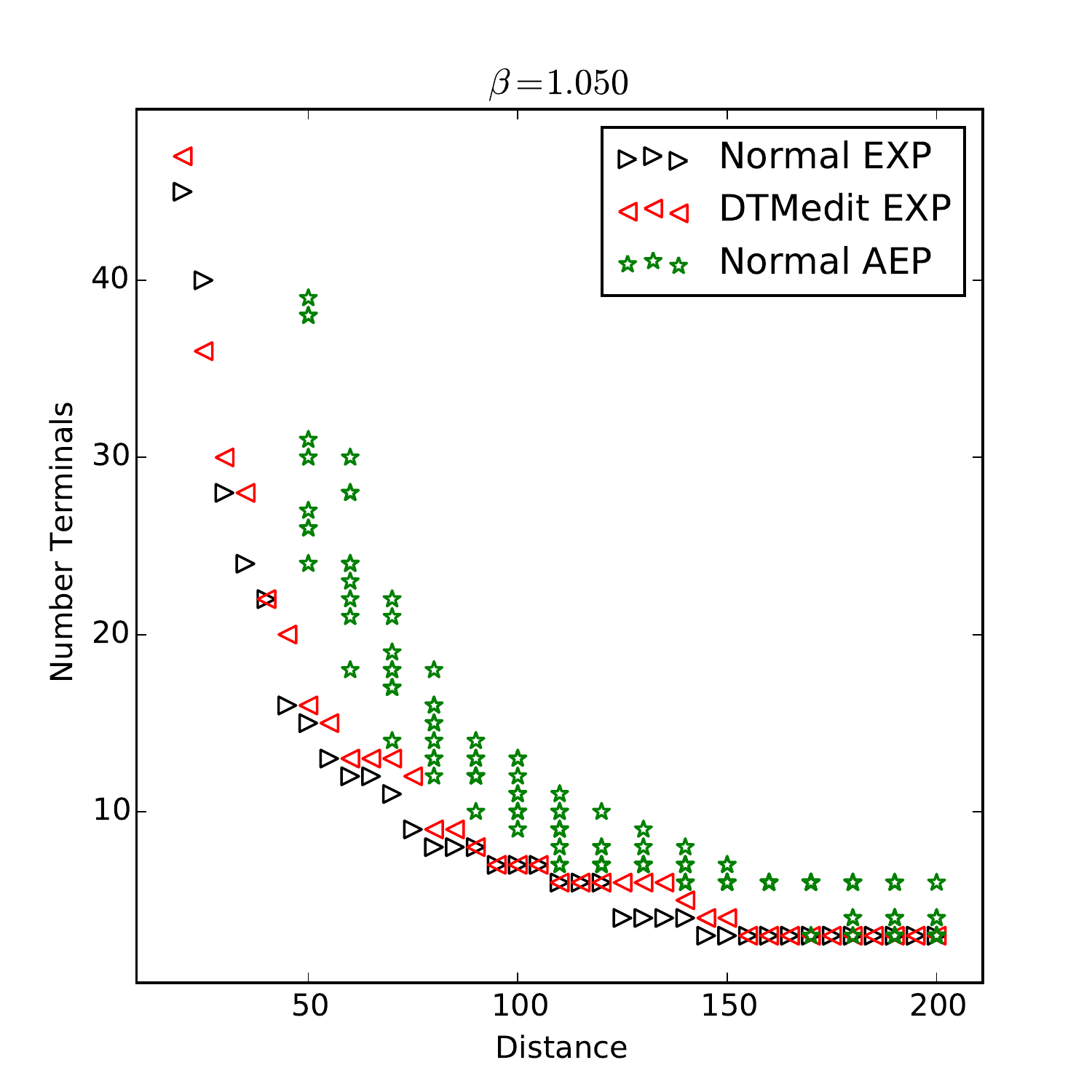}% This is a *.jpg file
 \end{center}
 \caption{Figure showing number of terminals for various parameters.  The black triangles pointing right are for the normal distance set while the red left pointing triangles are for the Delaunay Triangulation derived distances, both using the the exponential deterrence function. The green stars represent the normal distance used with the ariadne style deterrence function \tref{detfuncaep} using all nine possible combinations of $\alpha = 3.6$, $4.0$ or $4.4$ along with $\gamma=0.9$, $1.0$ or $1.1$. These are all for $\beta=1.05$.}\label{fPEXPAEPb1050DvsT}
\end{figure}

\begin{figure}[h!]
 \begin{center}
 % produced by terminalAnalysis.py
  \includegraphics[width=12cm]{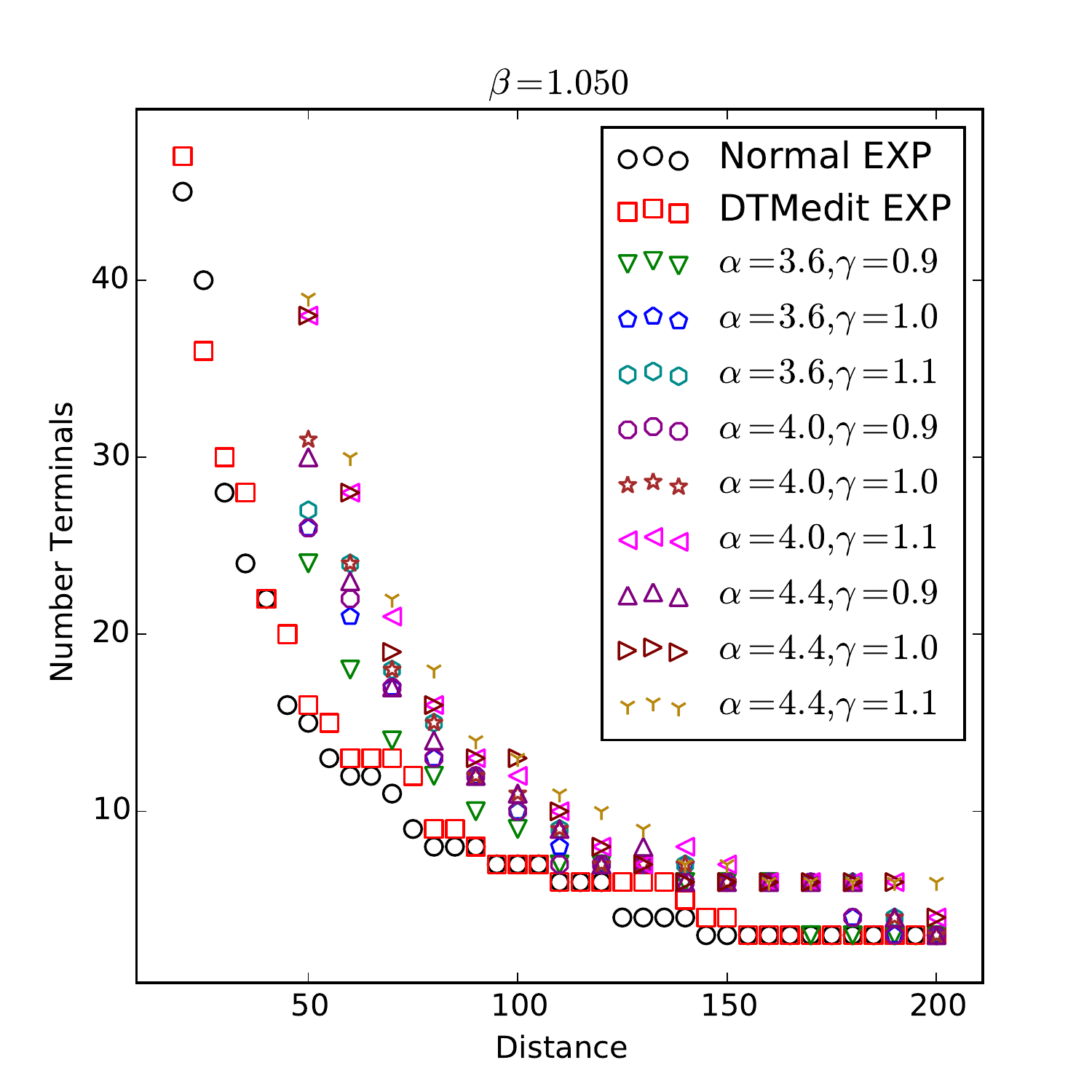}% This is a *.jpg file
 \end{center}
 \caption{Figure showing number of terminals for various parameters.  The black circles are for the normal distance set while the red squares are for the Delaunay Triangulation derived distances, both using the the exponential deterrence function. These are generally below the other points which are for the normal distance sets used with the ariadne style deterrence function \tref{detfuncaep} using all nine possible combinations of $\alpha = 3.6$, $4.0$ or $4.4$ along with $\gamma=0.9$, $1.0$ or $1.1$ as indicated by the legend. These are all for $\beta=1.05$.}\label{fPEXPAEPb1050DvsTag}
\end{figure}

Both here and in the original papers changing the non-linearity parameter $\beta$ in the range $1.05$ to $1.25$ does not seem to change the qualitative nature of the results for choices of $D$; the pattern of terminals is qualitatively the same.  So we will fix $\beta$ and then look at the remaining uncertainty in further detail.  We chose to fix $\beta=1.05$ as a typical value from the earlier studies. Initially we adopt the normal distance data set and the exponential deterrence function of \tref{detfuncexp} corresponding to the case where cost equals distance. The sensitivity to distance scale can be seen clearly for this fixed beta value in figs. \ref{fPEXPAEPb1050DvsT} and \ref{fPEXPAEPb1050DvsTag}.

We have also repeated the analysis for our Delaunay Triangulation based distances.
The change of the distances from normal to Delaunay Triangulation derived values has only a small effect.  Generally the latter reproduce the same number of terminals at a slightly higher distance scale. This is to be expected as the distances between any two points in the Delaunay derived distances is always equal to or greater than the same points in the normal distance set.  The main feature is that both these two cases using the exponential deterrence function show the instability in solutions with respect to small changes in the parameters as seen from the contour plots in \figref{fPEXPbvsDTcontour}.  So solutions with a given number of terminals $T$, when $T$ is bigger than three, are only produced for very limited ranges of $D$, typically less 10 or less, roughly the nearest neighbour site separation.

As a separate exercise Fig.s~\ref{fPEXPAEPb1050DvsT} and \ref{fPEXPAEPb1050DvsTag} illustrate the effects of changing the deterrence function to the {\it ariadne} form \tref{detfuncaep} with the values of its two further calibration parameters $\alpha$ and $\gamma$ varied by 20\%. We see the same instability as seen in the contour plots in terms of small changes in parameter $D$ until we get to large distance scales with six or less terminal sites in their solution.  Again the most stable solution has three terminals.  The most interesting effect is that when we compare the solutions with the same number of $T$ terminals and the same normal distance set, we see that changing the shape of the potential changes the value of the distance scale $D$.

This leads to an important conclusion. We should not try to compare two different theories at the same distance scale directly, even though in some qualitative way they correspond to the mode of transport.  The relationship between a parameter of a theory, here $D$, and the actual physical properties it corresponds to, perhaps the typical physical separation of our terminals, is highly non-trivial and depends on the details of the theory.  This is a simple example of what is called  `renormalisation' in physics. What we must do is to pick a physical property, say the number of terminals, and then find the values of the a theoretical parameter such as $D$ where different models give similar physical results.  Here the exponential potential distances scales can be as much as half the {\it ariadne} deterrence function distance scales for results with the same number of terminals.

So for the next stage of investigation we fix $\beta$ and find solutions with a specified number $T$ of terminal sites.

% ---------------------------------------------------------
\subsection{Eight terminals}\label{ss8terminals}

\begin{figure}[h!]
 \begin{center}
 % Figure produced from terminalAnalysis.py using data in terminalnumberb1050.dat derived from RW109bEither_RWGM_PEXPAEP_mastats.xlsx
  \includegraphics[width=8cm]{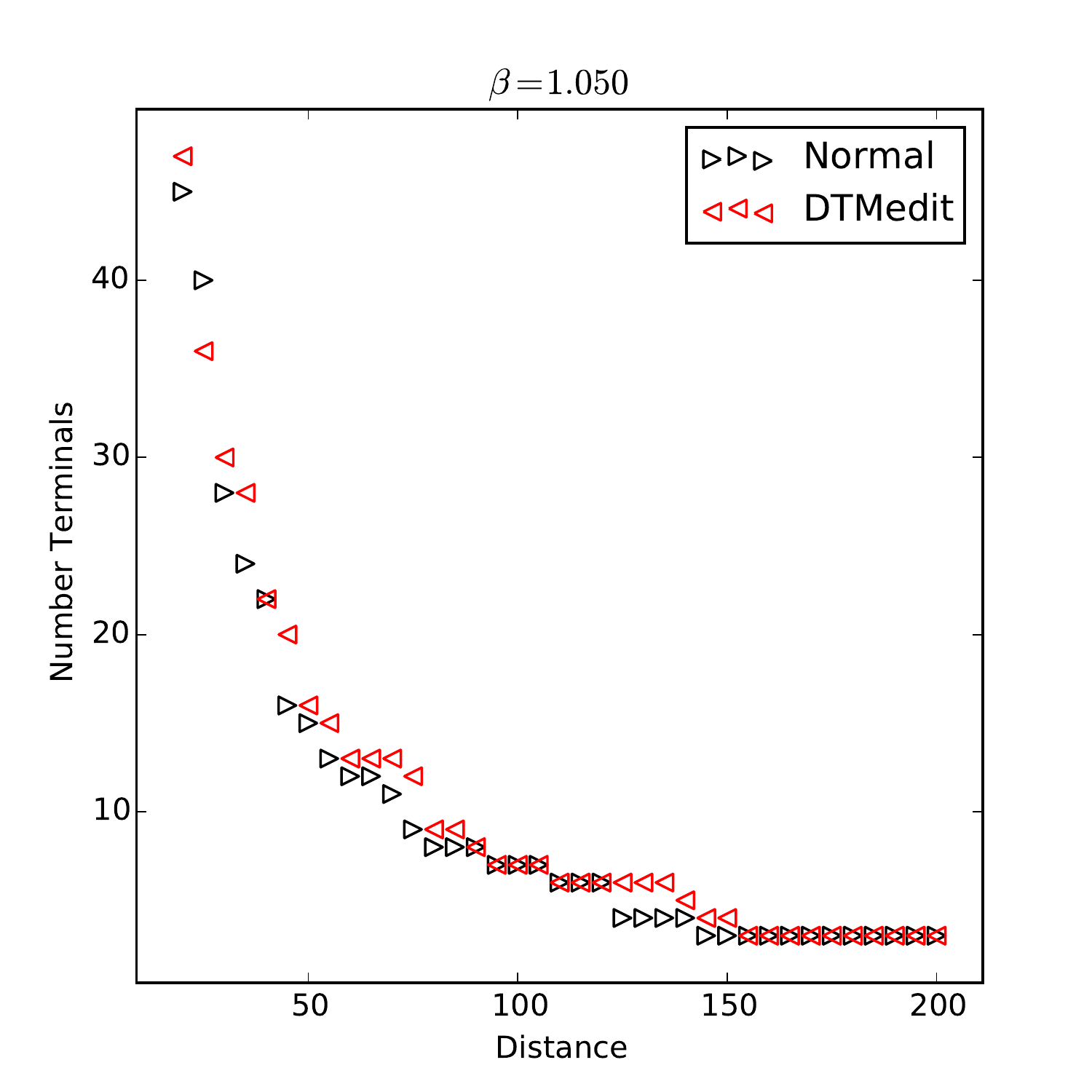}
 \end{center}
 \caption{Figure showing the different number of terminal sites found using exponential deterrent function with $\beta=1.05$. For normal (direct) distances and for distances based on modified Delaunay Triangulation (DTMedit).}\label{fRWGM_PEXP_b1050_DvsT}
\end{figure}

We choose to focus on model outputs with $\beta=1.05$ and eight terminals, to see to what extent we can replicate Rihll and Wilson and how to take the analysis further.  These are typical values considered in the original paper \citet{RW91}, with Fig.6  satisfying the criteria exactly.  To do this we start with our best representation of the distances between sites as outlines in section \ref{sssitedata}. We set $\beta=1.05$ and, initially, adopt the same exponential form for the deterrence function as shown in \tref{detfuncexp}. We have to find our own distance scale $D$ as the units used in the original papers are unknown.  To do this we scan through all possible value for $D$, other parameters fixed as described, looking for solutions with 8 terminals. We also repeated this exercise for the distances derived from our modified Delaunay Triangulation. The results are shown in \figref{fRWGM_PEXP_b1050_DvsT}.

We found three distinct solutions with eight terminals at $D=80$, $85$, and $90$ for our normal distance matrix and only one at $D=90$ for our modified Delaunay Triangulation (DTMedit) distance set. The precise sites are shown in Tables \ref{t8terminalsordered} and \ref{t8terminalsregion}.

\begin{table}
\begin{center}
\begin{tabular}{c||c|c|c|c|c}
Data Set & DTMedit & Normal & Normal & Normal & RW91  \\
         & $D=90$  & $D=90$	& $D=85$	 & $D=80$	  & Fig 6 \\ \hline \hline
Top Site  & Potniai 26	& Kabirion	& Kabirion	& Kabirion& Thebes (25) \\ \hline
2nd       & Medeon 17   &	Athens	 &	Onchestos	 &	Onchestos	 &	Akraiphnion (7)  \\ \hline
3rd       & Berbati 96  &		Prosymnia	 &	Athens	 &	Athens	 &	Koroneia (23) \\ \hline
4th       & Koropi 57   &		Koropi	 &	Prosymnia	 &	Koropi 57	 &	Athens \\ \hline
5th       & Athens 70	&	\begin{tabular}{@{}c@{}} Argive \\ Heraion 98 \end{tabular} 	 &	Koropi	 &	 Prosymnia 97	 &	 Argos (101)\\ \hline
6th       & Korinth 82	&	Onchestos	 &	\begin{tabular}{@{}c@{}} Argive \\ Heraion 98 \end{tabular}
                                                         &	\begin{tabular}{@{}c@{}} Argive \\ Heraion 98 \end{tabular} 	
                                                         &	Kalyvia 59\\ \hline
7th       & \begin{tabular}{@{}c@{}} Argive \\ Heraion 98 \end{tabular}
                       &		Kromna	 &	Kromna	 &	Kromna 78	 &	Korinth (82)\\ \hline
\begin{tabular}{@{}c@{}} Weakest \\ terminal \end{tabular} & Mykalessos 15	 &	Aulis &		 Aulis &		Aulis 14 &		 Khalkis (40)
\end{tabular}
\end{center}
\caption{The eight terminal sites found for $\beta=1.05$ using an exponential deterrence function, ordered by the strength of the terminal flow. One using our modified Delaunay Triangulation (DTMedit) derived distances, one with out direct distances (normal) and the last taken from fig.6 of \citet{RW91}.}\label{t8terminalsordered}
\end{table}

\begin{table}
\begin{center}
\begin{tabular}{c|c|c||c}

DTMedit & Normal           & RW91    & Region \\
$D=90$  & $D=80,85,90$	  & Fig 6   &        \\ \hline \hline
Mykalessos 15 &	Aulis 14     &	Khalkis 40	  &	Near Euboea \\ \hline
Potniai 26	  & Kabirion 24  &	Thebes 25	  &	Near Thebes \\ \hline
Medeon 17	  & Onchestos 9  &	\begin{tabular}{@{}c@{}} Akraiphnion 7\\ Koroneia 23 \end{tabular}
                                              &	N.\ Boeotia \\ \hline
Athens 70	 &	Athens 70	 &	Athens 70	      &	Athens \\ \hline
Koropi 57	 &	Koropi 57	 &	Kalyvia 59    & S.\ Attica \\ \hline
Korinth 82	 &	Kromna 78	 &	Korinth 82    & Near Corinth \\ \hline
\begin{tabular}{@{}c@{}} Berbati 96 ,  \\ Argive Heraion 98 \end{tabular}	 &	
\begin{tabular}{@{}c@{}} Prosymnia 97, \\ Argive Heraion 98 \end{tabular}	 	
                             & Argos (101)	  & Near Argos
\end{tabular}
\end{center}
\caption{The eight terminal sites found for $\beta=1.05$ using an exponential deterrence function, organised by location. One using our modified Delaunay Triangulation (DTMedit) derived distances, one with out direct distances (normal) and the last taken from fig.6 of \citet{RW91}.}\label{t8terminalsregion}
\end{table}

The results show a lot of consistency with variations on the scale of about 10km. That is, all our examples give Athens as one of the dominant sites, as Rihll and Wilson also found.  There is also at least one terminal site close to Argos and another close to Corinth, but often it is one of their close neighbours and not the sites which became dominant in later times.  However this is on a relatively small scale of roughly the average site separation, under 10km. This `error' is emphasised by the fact none of our results here gives historic Thebes as the dominant site, but rather we find one of its close neighbours. Looking at \tabref{t8terminalsordered} we see the order of the sites, as defined by the terminal flow \tref{termdef}, moves around between the different solutions on top of the actual again emphasising the level of robustness of the results.

Perhaps more interestingly, \tabref{t8terminalsregion} shows a clear difference between the three distance sets, the original (unknown) Rihll and Wilson distances, our normal direct distances and the modified Delaunay Triangulation (DTMedit) derived distances. Apart from Athens, the dominant site in each region is different in at least two and mostly in all three cases. So we see that different choices of distance measures do have an effect.  On the other hand the changes are small, on this nearest neighbour scale of a few kilometres. Uncertainty on this scale is to be expected given the uncertainty of other aspects of the modelling. It is almost certainly also no worse than the uncertainty coming from our lack of detailed knowledge of the actual terrain (geographical, political and social) of the period.

% ---------------------------------------------------------
\subsection{Three terminals}\label{ss3terminals}

The clustering suggests that three terminal solutions are the only ones which are in a noticeably stable
region of parameter space.  We are not suggesting that, in this period, there were only three significant sites. Our purpose is to understand the nature of the modelling process better when there is less ambiguity. The three sites found for the exponential deterrence function and the two different distance sets are given in \tabref{t3terminal}. As might be expected given the noticeable gap visible to the eye in the map of sites in \figref{fRW109numbers} there is one site for each of the three clear regions: Boeotia, Attica and Isthmus/Argolid.
\\
\\
For Boeotia, yet again Thebes itself is never a terminal, but the terminal site picked is always close to Thebes.  Unlike the results for eight terminals, this site is no longer always a nearest neighbour, e.g.\ Plataia is about 13km south of Thebes.
\\
\\
For Attica, we find that the Delaunay Triangulation distance data no longer picks Athens but instead picks an extremely close neighbour of Athens. The difference between the two data sets reflects the fact that they differ only on longer distance scales.  With three terminals, each site is the centre of attraction for about a third of the sites spread over a much larger distance so the differences between the two distance sets will be more important when we have fewer terminals.
\\
\\
The most interesting case is the terminal in the Isthmus/Argolid region, either Berbati or Tenea being chosen. These sites are somewhere in between the two neighbourhoods, that close to Argos and the second around Corinth, which provided Ishmus/Argolid region terminal sites when there were eight terminals in total.  What this result suggests is that to a first approximation, the Rihll and Wilson model splits the space into roughly equal area patches and with one terminal per region, the terminal lying close to the geometric centre of its region.

\begin{table}[htbp]
\begin{center}
\begin{tabular}{c|c||c|c|c}
\hline
\textbf{Distance Data} & \textbf{Distance} & \multicolumn{3}{c}{\textbf{Terminal Sites}} \\
   & \textbf{D} & \textbf{Boeotia} & \textbf{Attica} & \textbf{Isthmus/Argolid}    \\ \hline\hline
Normal  & 150                          & 24 Kabirion & 70 Athens    & 96 Berbati \\ \hline
Normal  & \begin{tabular}{@{}c@{}} 155, 160, 165,   \\ 170, 175, 180 \end{tabular}
                                       & 31 Eutresis & 70 Athens    & 96 Berbati \\ \hline
Normal  & 185                          & 31 Eutresis & 70 Athens    & 89 Tenea   \\ \hline
Normal  & 190,195,200                  & 36 Plataia  & 70 Athens    & 89 Tenea   \\ \hline
DTMedit & 155,160,170,175              & 26 Potniai  & 71 Kallithea & 96 Berbati \\ \hline
DTMedit & 180,190,200                  & 26 Potniai  & 71 Kallithea & 89 Tenea
\end{tabular}
\end{center}
\caption{The three terminal sites for normal and Delaunay Triangulation distance data with the exponential deterrence function when solutions have exactly three terminals.}
\label{t3terminal}
\end{table}

\begin{figure}[h!]
 \begin{center}
  \includegraphics[width=16cm]{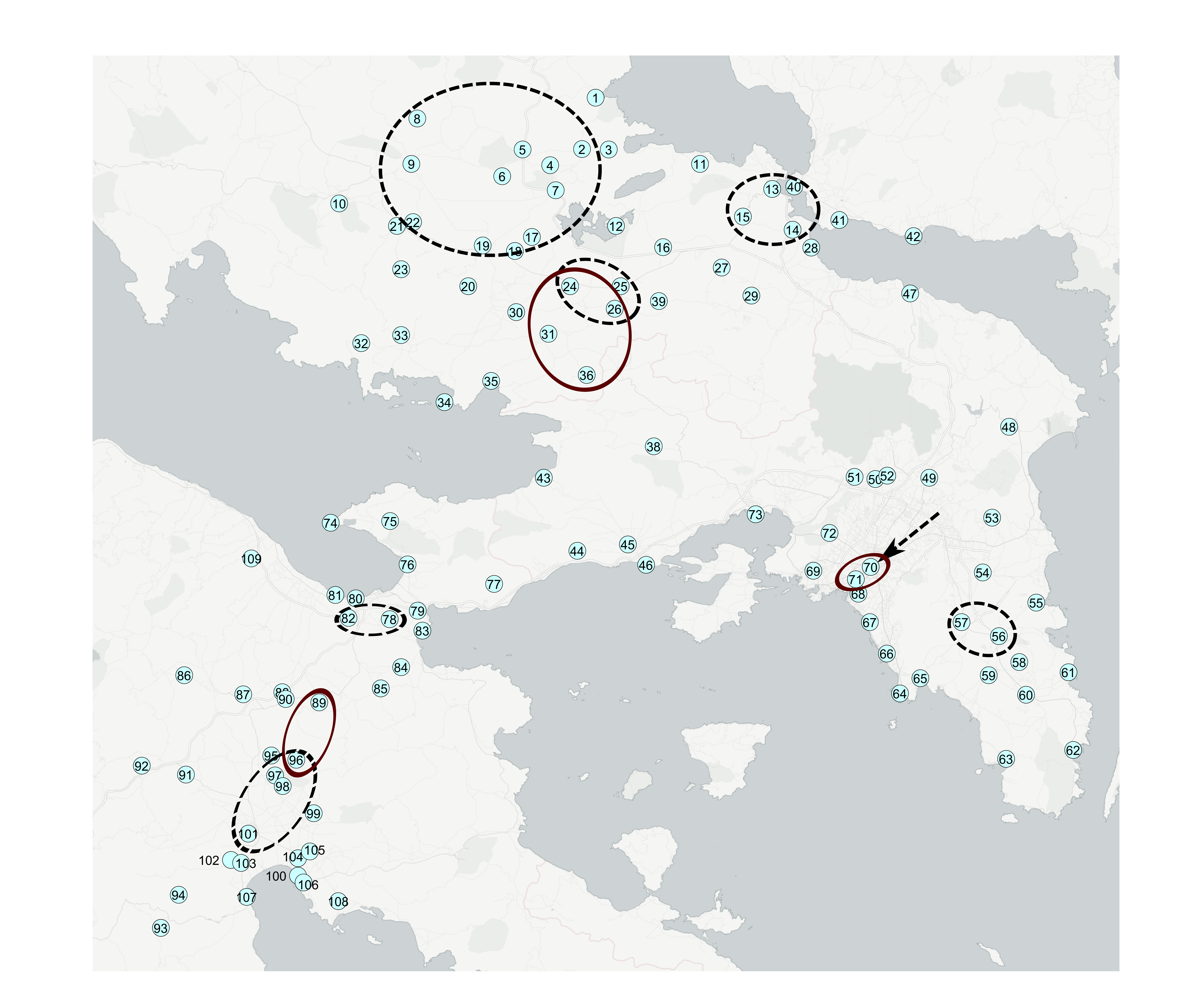}
 \end{center}
 \caption{Ranges of uncertainty for the terminal sites. The black dashed ellipses and the arrow indicate the range of possible locations for the eight terminal solutions of section \ref{ss8terminals}.  The red solid ellipses show where we find terminal sites when there are only three in total in any one solution as as discussed in section \ref{ss3terminals}.}\label{fRW109uncertainty}
\end{figure}

This has implications for our more general analysis, which we now test.

% ---------------------------------------------------------
\subsection{The Cluster-and-Centre method}

%\begin{itemize}
%\item Conjecture:- terminals are the closest to the geographical centre of the sites in their cluster.
%
%To answer this look at points on a line analytically.
%
%\item Conjecture:- if we clustered our distance data using a standard clustering method, and then found the site closest to the geographical centre of these clusters, these would be essentially the terminal sites found in RW model.
%
%\item To Do: k-means cluster the distance data. Find site closest to centre of cluster
%\end{itemize}

The typical type of network produced by the Rihll and Wilson model is shown in Fig.s \ref{fRWGMPEXPb1050d85} and  \ref{fRWGMPEXPb1050d90}. It shows a set of connected stars, zones of influence, where the terminal sites are each the centre of one star while remaining other sites only have one strong connection to a nearby terminal site. This effectively defines a clustering of the sites\footnote{This is called a \tdef{community} of sites in network science and, more formally, this is a \tdef{partition} of the set of sites.} where each cluster contains one terminal site along with all the sites with a strong connection to the terminal site of the cluster. Looking at the solutions it appears that the terminals are often close to the geometric centre of the clusters.  This is not surprising at the site closest to the centre is likely to minimise the sum over deterrence function terms in the denominator  of \tref{FRW} which would allow a larger $I_i$ value in the solutions and a large $I_j$ values characterises the terminal sites.

This leads us to the conjecture that the Rihll and Wilson model is clustering close sites and then giving the most central node as the terminal node.   One way to test this is to attempt a different approach to finding the key sites which we will refer to as a ``Cluster-and-Centre'' method.  This is a null phenomenological approach, relying on geographic data alone, without the trappings of entropy and Bayesian analysis. First we use a standard data clustering method, one which takes distances between data points (here the sites) as their input. The only parameter of the method is the number of clusters to be considered. Once we obtain our clusters of sites, we then find the most central site by looking for the site which is closest to the point with coordinates given by the mean (or the median) of the coordinates of all the sites in the cluster.

In \tabref{tclustercentre} we show the results for eight clusters so we can compare them with the eight terminal results discussed above. We used k-means clustering and five variations of Hierarchical Agglomerative Clustering (HAC) \citep[for example see][for both methods]{MRS08}. The implementation of k-means clustering that we use is stochastic so we are just showing one possible result here.

For three of the four flagship sites we are using to test of results, this approach gives reasonable results, except when using the method labelled `HAC single'.  The `single' version of Hierarchical Agglomerative Clustering is well known to be an extreme version and is often not appropriate. It is clearly an outlier here and we will ignore it from now on\footnote{Further discussion and more detailed results supporting this are given in the supplementary material, section \ref{appcandc}.}.
For the remaining five methods shown in \tabref{tclustercentre}, Argos is always picked as a centre.  Corinth is only picked out only once and either 80 Lekhaion or 78 Kromna are chosen instead. However these are close neighbours of Corinth and so we are finding the methods chooses the same type of 10km wide region as we did with the Rihll and Wilson model.  Athens is picked out most often in the northern Attica region (six times from the ten reasonable methods) with 57 Koropi or 50 Menidi picked otherwise. However neither of these alternative sites is particularly close to Athens, lying outside the range of results we found with the Rihll and Wilson model.

The big difference comes in the Boeotian region.  Thebes is never picked out. Further, the two close neighbours in the typical small region picked out in our analysis using the Rihll and Wilson model, 26 Potniai and 24 Kabirion, are only picked out in only four of the ten reasonable results.

Overall, while there is some correlation between the Rihll and Wilson model and the Cluster-and-Centre method outlined in this section, it is not an overwhelmingly strong or clear relationship. We need more than geography to understand why some sites develop to become so  important at the expense of their neighbours. The entropic/Bayesian approach of Rihll and Wilson provides a useful next step to our understanding of state formation.

% ************************************************************************
\section{Conclusions}

In our earlier paper \citep{RE14} we speculated on the contingent nature of Thebes in the Rihll and Wilson model, taking their results for granted. In this paper we have shown that Thebes (or, indeed, any site) can only be understood in the much larger context of uncertainty in spatial network modelling and how this is reflected in model outcomes, a context which has implications for all entropic attempts to understand urbanisation and city-state development.

We have highlighted several sources of uncertainty: site choice, distances, model choice, parameter choice. In this paper we have looked at the effect on outcomes of some (but not all) these sources of uncertainty.  We have tried two ways to calculate distances, our normal and Delaunay Triangulation derived distances.  By comparing with the results of Rihll and Wilson \citeyear{RW87,RW91} we have effectively a third set of distances. We have tried also two major ways to encode the costs of distance in the models: an exponential deterrence function \tref{detfuncexp} and the ariadne deterrence function \tref{detfuncaep}.

An important requirement of this work is that to evaluate uncertainties we must make fair comparisons between results from different variations of the same model or even completely different models.  The key problem is that we do not know what parameter values to use in different models in order to make this fair comparison. Even when a parameter is apparently linked to a physically measurable scale, such as our distance scale parameter $D$, the relationship between model value and actual measurable physical quantity is complicated.  Should we relate the the distance scale $D$ directly to the typical daily walking distance, or should that be $1.5D$ or $0.9D$? Whatever this relationship is, it will change with different model parameter values and indeed when using different models.  This is the key principle of `renormalisation', that we must never assume that a model parameter is simply related to a physical quantity.
The answer we suggest is to compare results from different models only when they are giving the same output by some suitable measure.

 In terms of the data used in this paper, we have arrived at similar conclusions to the original authors, though we put these within larger but quantified margins of uncertainty. Our results show that there are regions about 10km across which have distinct geographical advantages for encouraging urbanisation and which can help explain the different roles of sites in later periods. We do not predict that Thebes, and to a lesser extent Athens, always going to be these sites, unlike the original papers, but we do agree that sites in their close neighbourhood are continually shown to have this advantage.  Thebes is not necessary but something like it was always likely.

%Our point is that, given the level of uncertainty, pinpointing a single site as a future dominant city was unlikely to be a robust conclusion from modelling.  We do say that a slightly bigger but still small patch is compatible with all the uncertainties and the modelling can make a robust assertion about the advantages of such small patches.

Of course similar links have been made between geographical locations and the role of sites in history: Delos in the case of \citet{D82}, Knossos in the case of \citet{KER08}.  However in the latter case the modelling also suggested a that pair of candidates, Knossos and Malia, had these spatial advantages, again compatible with the uncertainties in modelling.

In terms of the method of Rihll and Wilson itself, our work has highlighted that, qualitatively at least, it seems to divide up the sites into regions of roughly equal size, each with a terminal site. To test the hypothesis that the terminals are close to the geometric centres of these zones,   we have compared this `zone-of-control' spatial ordering against a null `Cluster-and-Centre' method which is based on generic clustering methods using geographical data alone. Specifically, it takes any clustering method to create the clusters and then finds their geometric centres to locate the dominant site for each cluster. This enables us to come up put `error bars' on our results, estimates of the range of reasonable results as illustrated by our \figref{fRW109uncertainty}.

For the clustering methods we tried here, we found there is some correlation between the Rihll and Wilson model and the Cluster-and-Centre method. However it is not an overwhelmingly strong or clear relationship. It seems that we need more than geography alone, a need arguably satisfied in part by entropic/Bayesian analysis. It could be argued that the popular and generic data clustering methods we used here, k-means and Hierarchical Agglomerative Clustering, are more effective on higher dimensional data than our two-dimensional world-surface.   However, even if it had been in better agreement with our outputs, which we don't think is likely, part of the rationale for using the Rihll and Wilson model is the interpretation it brings.  The Rihll and Wilson model comes with a powerful epistemic interpretation in terms of the entropy of microstates and a natural interpretation in terms of flows, inputs, outputs and costs. We learn from the nature of the calibration. A regular data clustering method such as k-means brings no intrinsic interpretation to the table, it merely hopes to provide reasonable clusterings in a calibration-free way.

\section*{Conflict of Interest Statement}
%All financial, commercial or other relationships that might be perceived by the academic community as representing a potential conflict of interest must be disclosed. If no such relationship exists, authors will be asked to confirm the following statement:

The authors declare that the research was conducted in the absence of any commercial or financial relationships that could be construed as a potential conflict of interest.

\section*{Author Contributions}

This paper is based on a talk given by RJR at EAA, Glasgow 2015 and by TSE at Univ.\ Konstanz 20th September 2016 \citep{E16b}. TSE wrote the computer codes. TSE and RJR both contributed to all other aspects of the work.\tnote{The Author Contributions section is mandatory for all articles, including articles by sole authors. If an appropriate statement is not provided on submission, a standard one will be inserted during the production process. The Author Contributions statement must describe the contributions of individual authors referred to by their initials and, in doing so, all authors agree to be accountable for the content of the work. Please see  \href{http://home.frontiersin.org/about/author-guidelines}{Author and Contributors here} for full authorship criteria.}

%\section*{Funding}
%Details of all funding sources should be provided, including grant numbers if applicable. Please ensure to add all necessary funding information, as after publication this is no longer possible.

\section*{Acknowledgments}
RJR thanks the organisers of the session at EAA Glasgow where this work was presented. RJR and TSE would also thanks Ulrick Brandes and the Konstanz group for discussions and Visone \citep{BW04} support during the workshop on Archaeological Network Reconstruction.

\appendix
\renewcommand{\theequation}{A\arabic{equation}}
\section*{Supplementary Material}

% \href{http://home.frontiersin.org/about/author-guidelines}{Supplementary Material} should be uploaded separately on submission, if there are Supplementary Figures, please include the caption in the same file as the figure. LaTeX Supplementary Material templates can be found in the Frontiers LaTeX folder

% *****************************************************************************
\section{Index of Sites}\label{sindex}

A list of sites giving their index and name is in \tabref{tRW109index}. The positions are found from the locations of a digitised version of Fig.~1 in \citet{RW87} with site coordinates given in our units.  For instance in our `normal' distance set, which uses the direct distance between sites, we have that site 1, Laryma, and site 109, Sikyon are separated by a distance is $\sqrt{ (469.5 -190.5)^2 + (169.1-536.2)^2} \approx 461$ units. The actual straight line distance is about 82km so the distance scale used in this paper is roughly 6 units for 1 km.
\begin{table}[htbp]
\scriptsize
\begin{tabular}{cl|r|r||cl|r|r||cl|r|r}
1 & Larymna & 469.5 & 169.1 & 37 & Hysiai & 511.5 & 387.2 & 73 & Eleusis & 616.9 & 492.6 \\ \hline
2 & Ay.Ionnis & 435.8 & 200.4 & 38 & Erythrai & 553.5 & 374.1 & 74 & Akraia & 258.0 & 502.5 \\ \hline
3 & Meg.Katavothra & 459.7 & 207.0 & 39 & Skolos & 571.6 & 353.5 & 75 & Perakhora & 314.0 & 508.2 \\ \hline
4 & Ay.Marina & 426.7 & 220.2 & 40 & Khalkis & 645.7 & 234.2 & 76 & Loutraki & 324.7 & 535.4 \\ \hline
5 & Kopai & 404.5 & 217.7 & 41 & Lefkandi & 693.4 & 258.0 & 77 & Krommyon & 406.2 & 567.5 \\ \hline
6 & Olmous & 380.7 & 232.5 & 42 & Eretria & 746.1 & 269.5 & 78 & Kromna & 309.9 & 582.3 \\ \hline
7 & Akraiphnion & 441.6 & 235.8 & 43 & Pagai & 419.3 & 477.8 & 79 & Isthmia & 334.6 & 574.9 \\ \hline
8 & Aspledon & 338.7 & 203.7 & 44 & Tripodiscos & 466.3 & 521.4 & 80 & Lekhaion & 277.8 & 566.7 \\ \hline
9 & Orchomenos & 295.1 & 218.5 & 45 & Megara & 509.9 & 519.8 & 81 & Ay.Gerasimos & 261.3 & 564.2 \\ \hline
10 & Lebedea & 255.6 & 244.0 & 46 & Nisaia & 522.2 & 535.4 & 82 & Korinth & 273.7 & 584.8 \\ \hline
11 & Anthedon & 566.7 & 214.4 & 47 & Oropos & 752.7 & 316.5 & 83 & Kenchraia & 337.0 & 588.1 \\ \hline
12 & Schoinos & 477.8 & 251.4 & 48 & Marathon & 852.3 & 440.7 & 84 & Solygeia & 325.5 & 615.2 \\ \hline
13 & Hyria? & 625.9 & 239.1 & 49 & Kephisia & 772.4 & 470.4 & 85 & Athikia & 300.8 & 640.7 \\ \hline
14 & Aulis & 636.6 & 254.7 & 50 & Menidi & 725.5 & 476.1 & 86 & Philious & 135.4 & 624.3 \\ \hline
15 & Mykalessos & 602.9 & 258.8 & 51 & Loissia & 697.5 & 463.8 & 87 & Nemea & 179.0 & 639.9 \\ \hline
16 & Glisas & 533.7 & 277.8 & 52 & Koukouvaones & 738.7 & 467.9 & 88 & Kleonai & 216.9 & 634.2 \\ \hline
17 & Medeon & 411.1 & 279.4 & 53 & Draphi & 836.6 & 501.6 & 89 & Tenea & 259.7 & 651.4 \\ \hline
18 & Onchestos & 396.3 & 296.7 & 54 & Spata & 801.2 & 544.4 & 90 & Zygouries & 231.7 & 649.0 \\ \hline
19 & Haliartos & 373.3 & 287.7 & 55 & Brauron & 856.4 & 566.7 & 91 & Sch.Melissi & 142.8 & 719.8 \\ \hline
20 & Askra & 365.8 & 323.9 & 56 & Markopoulo & 841.6 & 574.1 & 92 & Orneai & 100.0 & 664.6 \\ \hline
21 & Itonion & 288.5 & 287.7 & 57 & Koropi & 800.4 & 579.8 & 93 & Hysiai & 110.7 & 830.9 \\ \hline
22 & Alalkomenai & 309.9 & 286.0 & 58 & Merenda & 821.8 & 593.0 & 94 & Kenkhraia & 131.3 & 808.6 \\ \hline
23 & Koroneia & 289.3 & 300.0 & 59 & Kalyvia & 819.3 & 622.6 & 95 & Mykenai & 207.0 & 688.5 \\ \hline
24 & Kabirion & 455.6 & 319.8 & 60 & Keratea & 867.1 & 639.9 & 96 & Berbati & 235.0 & 697.5 \\ \hline
25 & Thebes & 487.7 & 319.8 & 61 & Kaki\_Thalassa & 900.0 & 658.0 & 97 & Prosymnia & 211.9 & 707.4 \\ \hline
26 & Potniai & 489.3 & 337.9 & 62 & Thorikos & 885.2 & 681.1 & 98 & Argive\_Heraion & 218.5 & 717.3 \\ \hline
27 & Eleon & 577.4 & 300.8 & 63 & Anavysos & 835.0 & 692.6 & 99 & Dendra & 251.4 & 740.3 \\ \hline
28 & Dramesi & 656.4 & 281.9 & 64 & Vouliagmeni & 742.0 & 639.1 & 100 & Pronaia & 233.3 & 788.1 \\ \hline
29 & Tanagra & 640.7 & 323.9 & 65 & Vari & 763.4 & 628.4 & 101 & Argos & 193.0 & 744.4 \\ \hline
30 & Thespiai & 405.3 & 349.4 & 66 & Aliki & 728.8 & 607.0 & 102 & Kephalari & 160.9 & 766.7 \\ \hline
31 & Eutresis & 434.2 & 351.9 & 67 & Trachones & 719.8 & 574.9 & 103 & Magoula & 177.4 & 774.1 \\ \hline
32 & Khorsia & 263.0 & 370.8 & 68 & Phaleron & 695.9 & 565.0 & 104 & Tiryns & 235.0 & 774.9 \\ \hline
33 & Thisbe & 300.8 & 363.4 & 69 & Kokkinia & 658.0 & 542.0 & 105 & Prof.Elias & 251.4 & 768.3 \\ \hline
34 & Siphai & 351.9 & 405.3 & 70 & Athens & 716.5 & 532.1 & 106 & Nauplia & 229.2 & 794.7 \\ \hline
35 & Kreusis & 378.2 & 393.8 & 71 & Kallithea & 707.4 & 542.0 & 107 & Lerna & 186.4 & 807.0 \\ \hline
36 & Plataia & 466.3 & 388.9 & 72 & Aigelaos & 673.7 & 509.9 & 108 & Asine & 267.9 & 817.7 \\ \hline
   &         &       &       &    &          &       &       & 109 & Sikyon & 190.5 & 536.2
\end{tabular}
\caption{Index of 109 sites used here. The coordinates are based on the digitisation of Fig.~1 of \citet{RW87}. The distances in this paper are based on these coordinates. For more details see \citet{E16a}.}
\label{tRW109index}
\end{table}

%\begin{figure}[h!]
% \begin{center}
%  \includegraphics[width=12cm]{RW87fig1screenshot}% This is a *.jpg file
% \end{center}
% \caption{Fig.~1 of \citet{RW87}.  This is used to define the locations of the 109 sites used as the starting point for this study. Note that site 64, Vouliagmeni, is not labelled on the figure, see \citet{E16a} for more details. \textbf{*** Need to get permission to use this figure. Need to replace with our own geographical figure. ***}}\label{fRW87fig1}
%\end{figure}

\begin{figure}[h!]
 \begin{center}
  \includegraphics[width=16cm]{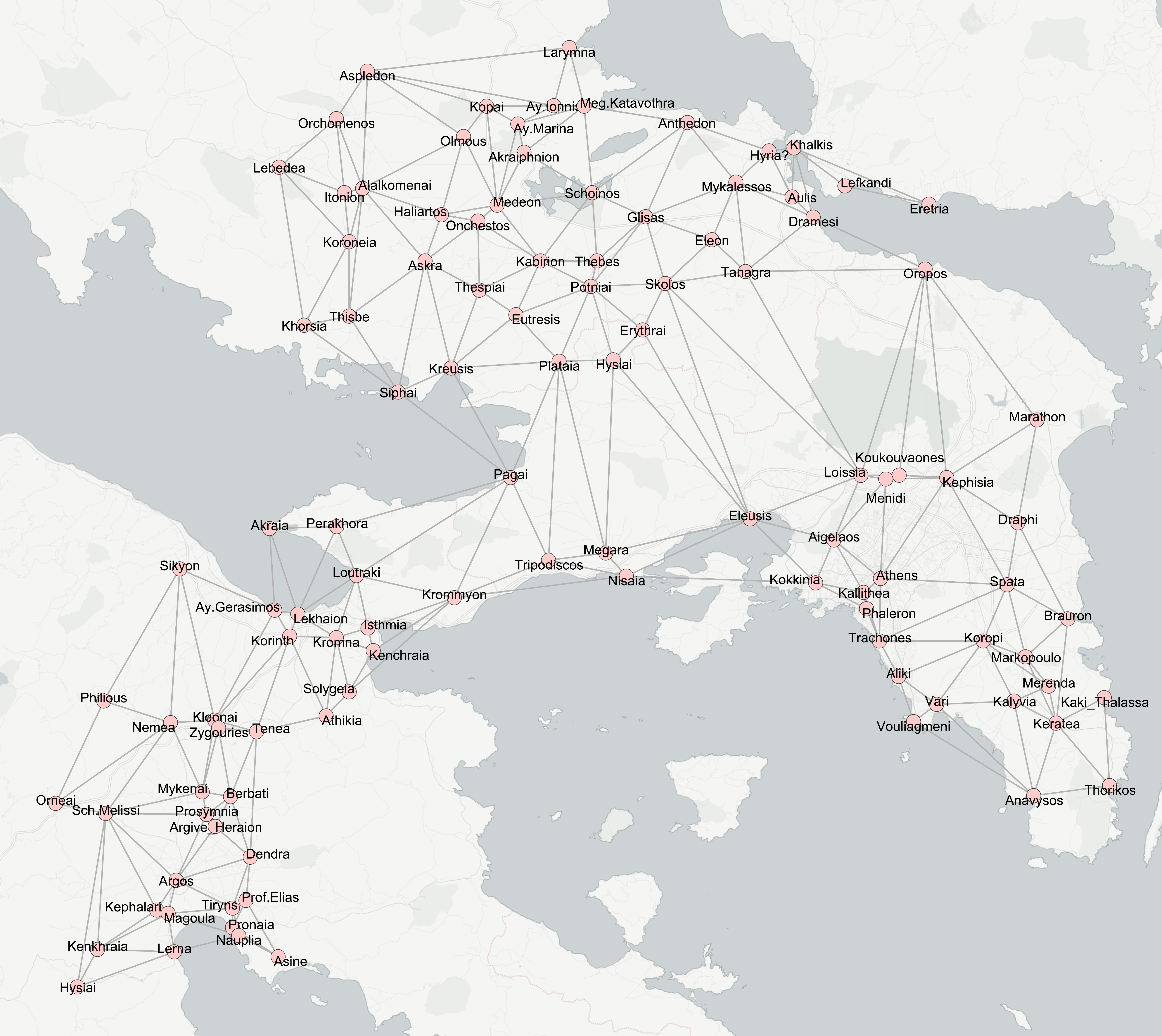}% This is a *.jpg file
 \end{center}
 \caption{The approximate locations of the 109 sites used as the starting point for this study, derived from Fig.~1 of \cite{RW87}. The index of site numbers is given in \tabref{tRW109index} in the Supplementary Material. Note that site 64, Vouliagmeni, was not labelled in the original figure, see \citet{E16a} for more details. The edges are those used to derive the second set of distances (denoted DTMedit) and are a subset of the edges of a Delaunay Triangulation.}\label{fRW109names}
\end{figure}

\begin{figure}[h!]
 \begin{center}
  %Figure showing HAC for RW109b and RW109bDTMedit
  %\includegraphics[width=16cm]{RW109baverage}% This is a *.jpg file
  \includegraphics[width=16cm]{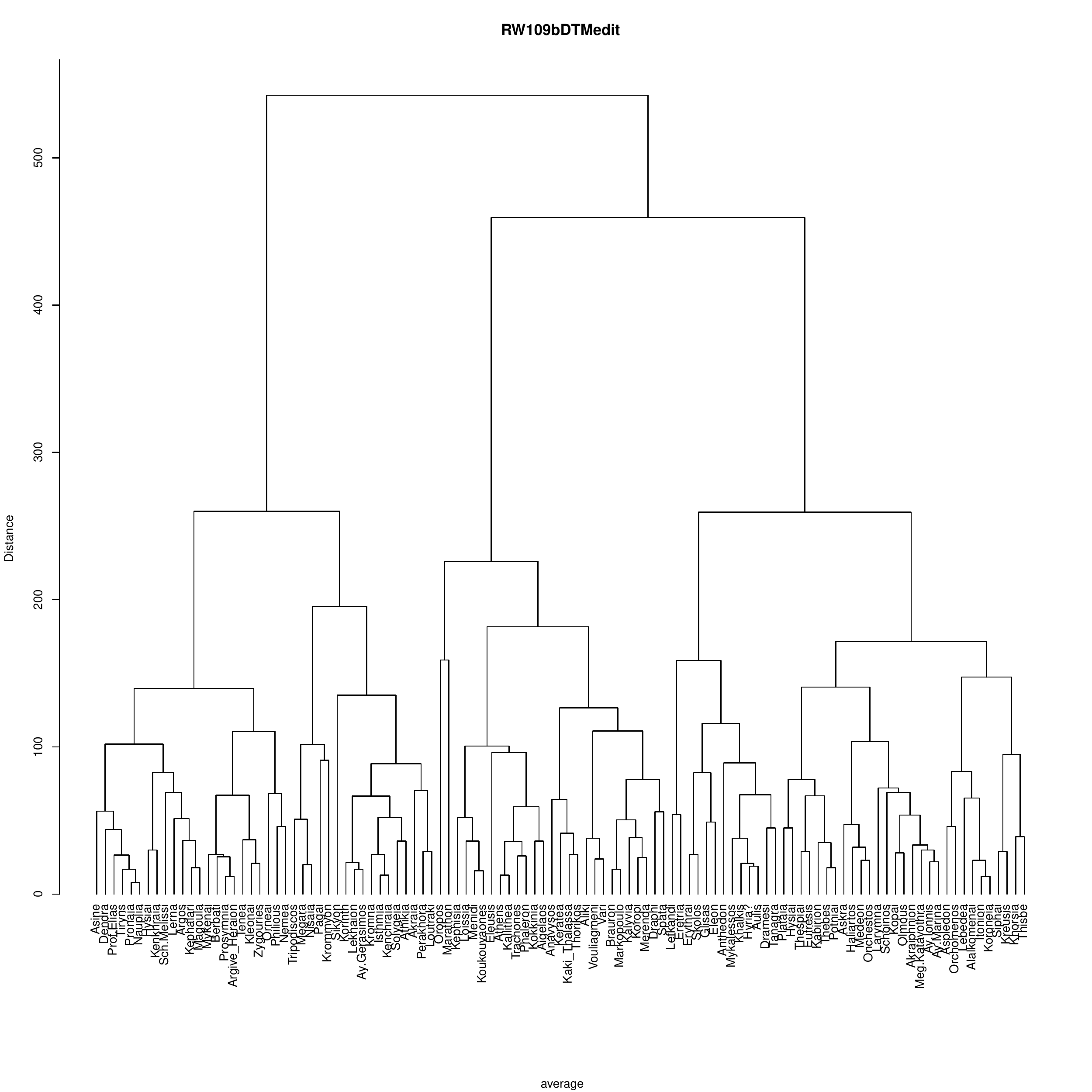}% This is a *.jpg file
 \end{center}
 \caption{Hierarchical Agglomerative Clustering for distances based on the edited Delaunay Triangulation network (DTMedit) using average criterion for agglomeration.}\label{fhacDTMedit}
\end{figure}

% ************************************************************
\section{Technical Details of the Rihll and Wilson Gravity model}\label{sRWtech}

The model used in the work of \citeauthor{RW87} \citeyearpar{RW87,RW91} was originally devised to determine the best position of retail centres \citep{H64,LH65,HW78} (see p.11 of \citet{RW87} for further citations).

Consider a network of $N$ sites labelled $i= 1,2,... ,N$. The model produces the link strengths $F_{ij}$ which describe the flows from sites $i$ to  sites $j$.  As a result the total \emph{outflow} from the site $i$ is $O_i = \sum_j F_{ij}$ and the total \emph{incoming} flow to site $j$ ps $I_j = \sum_k F_{kj}$. Here we will outline the derivation of the model using an entropy viewpoint pioneered by Wilson in the context of generic spatial modelling \citep{W67}.  The key assumption is that if all other things are equal, then every possible exchange counted by the flows is equally likely.  Put another way, if there is no other information about these exchanges then the best we can do is to assume all exchanges are equally likely. The maximum entropy framework provides a rigourous mathematical basis for this simple idea.

Of course in reality there are strong constraints and different models add in different types of extra information in an attempt to provide a more realistic description of the individual exchanges. Equivalently, they can be understood as the most likely configurations of a microcanonical ensemble subject to these same constraints.

For the shopping model \citep{H64,LH65,HW78} used by Rihll and Wilson, the flows defined by the model maximise the entropy $S$, more conveniently understood as minimising the Hamiltonian $H = -S$ where, after some rewriting, \cite[see][equation 30-33]{HW78}
\bea
 H &=& \sum_{i,j} F_{ij} \big(\ln(F_{ij})  -1  \big) -\sum_i \alpha_i  \Big[O_i-\sum_j F_{ij} \Big] -\beta \Big[C-\sum_{i,j}(F_{ij}c_{ij})\Big]
 \nonumber \\
 && \hspace*{5cm} + \beta \Big[X-\sum_{j}I_j \big(\ln I_j -1\big) \Big],
 \label{HdefRW2}
\eea
The solutions are the most likely pattern for exchanges given the constraints (given in the square brackets in \tref{HdefRW2}) imposed in the model. For simplicity we are assuming that every site $i$ has the same number of potential origins or destinations for a trip.
The first term reflects the assumption that the probability that an exchange occurs on \emph{any} given edge is independent of the edge if all other things are equal. The second term, with coefficient $\alpha$ imposes a constraint that the total output of site $i$, the total flow leaving site $i$, is equal to a parameter $O_i$ which we must specify.  The third term supposes that the cost of each exchange from $i$ to $j$ is $c_{ij}$, and that we demand that the total cost is $C$, another parameter we must satisfy.  This `cost' is not necessarily in terms of money.  Rather it is generally expressed in terms of some characteristic fixed by the geography of the site, typically some measure of the distance between two sites.  All of the first three terms are frequently seen in this type of entropy approach. It is the last term which is distinctive and is a key feature of the model of Rihll and Wilson. It involves the total \emph{incoming} flow, $I_j = \sum_k F_{kj}$, but it is not a simple constraint on the total input to each site.

 Rather, it is a constraint on the entropy of site outflows which, in the absence of this (and other constraints) would be uniform. To understand what type of effect this last term is giving, and indeed to get a better understanding of what all the terms lead to, it is easier to quote the solution for the pattern of flows which maximises the entropy, which can be expressed in simple algebraic form as
\beq
 F_{ij}=A_i O_i I_j^\beta f_{ij} \, , \qquad  A_i^{-1} = \sum_k I_k^\beta f_{ik} \qquad  I_j = \sum_i f_{ij} \, .
\label{FRWapp}
\eeq
The parameters $\alpha_i$ have been fixed by the requirement that the outputs for each site are fixed to be are $O_i$ from which the normalisation factors $A_i$ are determined in terms of other quantities in the theory. The total cost $C$ and the cost of each exchange $c_{ij}$ are equivalent to specifying what is known as the {\it deterrence function} $f_{ij}$,
\beq
 f_{ij} = \exp(- c_{ij}/D)
 \label{detfuncdef}
\eeq
where  $D$ is a distance scale against which `costs' are measured. More generally, we write
\beq
 f_{ij} = f(d_{ij}/D)
 \label{detfuncexp1}
\eeq
where $f(x)$ relates costs to the separation variables $d_{ij}$ between sites $i$ and $j$. For instance if the costs are just the distances, $c_{ij}=d_{ij}$ then we arrive at a simple exponential form
\beq
 f_{ij} = \exp(-d_{ij}/D )
 \label{detfuncexp2}
\eeq

We stress that the outflows $O_i$ are \emph{input} parameters for the Rihll and Wilson model \tref{FRWapp} whereas the inflows
$I_j= \sum_i F_{ij}$ are to be determined from \tref{FRWapp} as \emph{outputs} from the model. The inflows are interpreted as the attractiveness or importance of a site and as used to determine the dominant city state sites.

In this case varying site size can be accommodated.
The obvious extension to include site size (that has its counterpart in the work of \citet{A78}) is to take
\bea
 H &=&  \sum_{(i,j)} F_{ij} \Big[\ln\Big(\frac{F_{ij}}{S_iS_j}\Big)  -\sum_i \alpha_i  \Big[O_i-\sum_j F_{ij} \Big] -\beta \Big[C-\sum_{i,j}(F_{ij}c_{ij})\Big]
 \nonumber \\
 && \hspace*{5cm}- \beta \Big[X-\sum_{j}I_j \Big(\ln \Big(\frac{I_j}{S_j}\Big) -1\Big) \Big],
 \label{HdefRW3}
\eea
Extremal solutions now take the form
\beq
 F_{ij} = S_i S_j^{1-\beta}e^{-\alpha_i} I_j^{\beta} f_{ij}.
\eeq
Consistency then requires that
\beq
F_{ij}=A_i O_i S_i^{1-\beta}I_j^\beta f_{ij} \, , \qquad  A_i^{-1} = \sum_k S_j^{1-\beta} I_k^\beta f_{ik} \, .
\label{FRW2app}
\eeq
We said that, in the absence of further evidence, we had taken all the $S_i$ to be identical. We note that explicit dependence on site size is very weak, going as $S_i^{1-\beta}$, where $1-\beta\approx 0.05$, with negligible effect. However, there is implicit dependence in the way the given outputs depend on site size (e.g. $O_i\propto S_i$).
% ------------------------------------------------------
\subsection{Non Linearity in the Rihl and Wilson Model}

The factor of $I_j^\beta$ in the Rihll and Wilson model is the key difference from most other gravity models. This term introduces non-linear feedback based on the inputs though the self-consistent normalisation factors.

To understand how this works, we will consider a simple way to find the solution to \tref{FRWapp} for a given set of parameters. The idea is that if we are given a set of input flows at some time $t$, say $I_j(t)$, then we specify the next round of values at iteration number $(t+1)$, $I_j(t+1)$.  To show this iterative process, we first rewrite the solution \tref{FRW2app} in terms of just the total inputs $I_j$ for each site along with the other fixed parameters as (see equation~(23) of \citet{RW87} and appendix of \citet{RW91})
\beq
 I_{j} = \left(\sum_i  \frac{O_i f_{ij}}{ \sum_k I_k^\beta f_{ik}  } \right) I_j^\beta    \, .
 \label{Isol}
\eeq
Note that this solution shows explicitly that the model can be derived from the $N$ different $I_j$ values alone.
This is a non-linear equation which can be solved using standard methods.

In practice in this paper we used a simple approach where
at each step of the numerical process we have our current best guess for the input values $I_j(t)$. The next set of values, $I_j(t+1)$, is then defined to be
\beq
 I_{j}(t+1) = \left(\sum_i  \frac{O_i f_{ij}}{ \sum_k I_k^\beta f_{ik}  } \right) I_j^\beta (t)    \, .
 \label{Isolnum}
\eeq
We start the process using the site output values as initial values $I(t=0)=O_i$.
We iterate many times until the changes in the $I_j(t)$ values are small, as defined by us. For more details on the convergence and uniqueness of this approach see \citet{HW78}.

This iterative form \tref{Isolnum} is also useful as it helps us understand the role of the distinctive non-linear factors $I_j^\beta$ found in the Rihll and Wilson model. If one site, say $T$, has a large input flow, so $I_T(t) \gg I_j(t)$ ($j \neq T$), it is very ``attractive'' in the language of Rihll and Wilson.  The normalisation factor, the denominator in \tref{Isolnum}, will then be be dominated by the one large factor of $I_T(t)$.  This will pull down all values of $I_j(t+1)$ except for $I_T(t+1)$ which is the only one boosted by a large factor in the numerator, the $(I_T(t))^\beta$. This creates a feedback loop as at each stage the $I_T$ entry gets larger and the others smaller, reinforcing the process. The feedback is enhanced the larger $\beta$ is. The logical end is to have most $I_j$ becoming zero so that all the input goes to one site, $I_T$.

In practice the solutions are a little more complicated.  If we assume that the deterrence function becomes negligible for sites much more than distance scale $D$ apart, then a site with a growing attractiveness $I_T$ will only suppress the attractiveness of sites within a radius of $D$ or so. Basically we should expect to see the system split up into patches of radius $D$ , each with one dominant site.  This is roughly what is normally seen.  We have a pattern of stars where all the flow from most sites, $O_i$, is directed to just one site, the {\it terminal site} in their neighbourhood.  The formal definition of a terminal site used by Rihll and Wilson is a particular implementation of a scheme of \citet{ND61} and we will follow suit. We define terminal sites $T \in \Tcal \subset \Vcal$ to be sites where the total flow into a site is greater than the largest single flow out of that site along any edge
\beq
 T \in \Tcal
 \quad \mathrm{iff} \quad  I_T > F_{Tj} \quad  \forall j   \, .
 \label{termdef}
\eeq
Basically for a terminal site more flow comes in than leaves the site along any one edge. In practice for many of the parameter values found here we found that terminal sites were the only ones with any significant flows into their sites so the in-strength, $I_i$, is usually sufficient to detect these important sites.

% **********************************************************************
\section{Cluster-and-Centre Methods and Results}\label{appcandc}

The Cluster-and-Centre algorithm first takes the distances between the sites and uses these to produce a partition of the sites using a standard data clustering method.  Let us suppose that these clusters are $\{ \Pcal_c \}$ where every site in is one and only one cluster, i.e.\ $\bigcup_{c} \Pcal_c = \Vcal$, and $\Pcal_c \cap \Pcal_d = \empty$ if $c \neq d$. We then define the centre of each cluster $c$ to be at $(x_c,y_c)$. We will define the centre in two ways, using the mean
\beq
 x_c = \frac{1}{|\Pcal_c|} \sum_{i \in \Pcal_c} x_i \, ,
 y_c = \frac{1}{|\Pcal_c|} \sum_{i \in \Pcal_c} y_i \, ,
\eeq
or the median of the coordinates $(x_i,y_i)$ of sites $i$ in the cluster $c$, $i \in \Pcal_c$. Finally we define the {\it central site} to be the site closest to the central point.

The six clustering methods we have used are k-means and five variations of Hierarchical Agglomerative Clustering (HAC); for example see \citet{MRS08} for both methods. Hierarchical Agglomerative Clustering methods work by starting with each site in a group by itself.  A distance scale $\Dcal$ is slowly increased and two groups are joined together as soon as the distance between these two groups equals the scale $\Dcal$. The different types of Hierarchical Agglomerative Clustering differ in the way they define the distance between two groups of sites. For instance the single (or minimum) Hierarchical Agglomerative Clustering method joins two groups if just one pair of sites, one from each group, is separated by $\Dcal$, while the complete (or maximum) method only joins groups when all pairs of sites (one from each group) are separated by at least $\Dcal$.
The average Hierarchical Agglomerative Clustering method joins two groups together if the dendrogram scale $\Dcal$ is equal to the average distance between all possible pairs of sites, one site from each of the two groups.

These central sites of the Cluster-and-Centre algorithm can then be compared to terminal sites \tref{termdef} defined in the Rihll and Wilson model. More generally the partitions $\{\Pcal_c\}$ of the Cluster-and-Centre algorithm can be compared to clusters $\{C_c\}$ formed in the Rihll and Wilson model which we define as follows.  Let $T \in \Tcal$ be a terminal site as defined by \tref{termdef}.  Then we define a cluster $\Ccal_T$ for each terminal site $T$ where
\beq
 u \in \Ccal_T \quad \mathrm{iff} \quad F_{uT} =\max ( F_{uv} \, \forall \, v \in \Vcal) \, .
\label{RWpartition}
\eeq
Provided the largest flow leaving a non-terminal site is unique for each site, then this defines a partition, that is each site is in one and only one cluster, i.e.\  $\bigcup_{T \in \Tcal} \Ccal_T = \Vcal$, and $\Ccal_S \cap \Ccal_T = \empty$ if $S \neq T$.

In \tabref{tclustercentre} below, we show the results of several attempts to cluster the data using standard methods followed by finding the geometric centre of each cluster. We note that the clusters are much more uneven in size for the Hierarchical Agglomerative Clustering methods with the single method being particularly poor.  That is one reason we do not consider it seriously in the discussions in the main text but in any case this single method is known to produce long thin clusters which don't seem appropriate for this context.
\begin{table}[htbp]
\begin{center}
\scriptsize
\begin{tabular}{l||c|rl|rl}
Method & No.\ in Cluster & \multicolumn{2}{c|}{Centre by median} & \multicolumn{2}{c}{Centre by median}
 \\ \hline \hline
k-means & 14 & 7 & Akraiphnion & 7 & Akraiphnion \\ \hline
k-means & 10 & 22 & Alalkomenai & 21 & Itonion \\ \hline
k-means & 12 & 28 & Dramesi & 14 & Aulis \\ \hline
k-means & 11 & 36 & Plataia & 36 & Plataia \\ \hline
k-means & 13 & 59 & Kalyvia & 59 & Kalyvia \\ \hline
k-means & 13 & 70 & Athens & 50 & Menidi \\ \hline
k-means & 16 & 82 & Korinth & 80 & Lekhaion \\ \hline
k-means & 20 & 101 & Argos & 101 & Argos \\ \hline \hline
HAC average & 22 & 19 & Haliartos & 19 & Haliartos \\ \hline
HAC average & 11 & 26 & Potniai & 26 & Potniai \\ \hline
HAC average & 10 & 28 & Dramesi & 14 & Aulis \\ \hline
HAC average & 5 & 44 & Tripodiscos & 44 & Tripodiscos \\ \hline
HAC average & 1 & 48 & Marathon & 48 & Marathon \\ \hline
HAC average & 25 & 57 & Koropi & 57 & Koropi \\ \hline
HAC average & 12 & 80 & Lekhaion & 78 & Kromna \\ \hline
HAC average & 23 & 101 & Argos & 101 & Argos \\ \hline \hline
HAC complete & 10 & 23 & Koroneia & 23 & Koroneia \\ \hline
HAC complete & 23 & 24 & Kabirion & 24 & Kabirion \\ \hline
HAC complete & 10 & 28 & Dramesi & 14 & Aulis \\ \hline
HAC complete & 5 & 44 & Tripodiscos & 44 & Tripodiscos \\ \hline
HAC complete & 9 & 60 & Keratea & 59 & Kalyvia \\ \hline
HAC complete & 17 & 70 & Athens & 70 & Athens \\ \hline
HAC complete & 12 & 80 & Lekhaion & 78 & Kromna \\ \hline
HAC complete & 23 & 101 & Argos & 101 & Argos \\ \hline \hline
HAC median & 27 & 16 & Glisas & 12 & Schoinos \\ \hline
HAC median & 14 & 22 & Alalkomenai & 22 & Alalkomenai \\ \hline
HAC median & 2 & 42 & Eretria & 42 & Eretria \\ \hline
HAC median & 5 & 44 & Tripodiscos & 44 & Tripodiscos \\ \hline
HAC median & 13 & 59 & Kalyvia & 59 & Kalyvia \\ \hline
HAC median & 13 & 70 & Athens & 50 & Menidi \\ \hline
HAC median & 12 & 80 & Lekhaion & 78 & Kromna \\ \hline
HAC median & 23 & 101 & Argos & 101 & Argos \\ \hline \hline
HAC ward & 19 & 17 & Medeon & 17 & Medeon \\ \hline
HAC ward & 10 & 23 & Koroneia & 23 & Koroneia \\ \hline
HAC ward & 14 & 28 & Dramesi & 14 & Aulis \\ \hline
HAC ward & 5 & 44 & Tripodiscos & 44 & Tripodiscos \\ \hline
HAC ward & 12 & 59 & Kalyvia & 59 & Kalyvia \\ \hline
HAC ward & 14 & 70 & Athens & 70 & Athens \\ \hline
HAC ward & 12 & 80 & Lekhaion & 78 & Kromna \\ \hline
HAC ward & 23 & 101 & Argos & 101 & Argos \\ \hline \hline
HAC single & 41 & 12 & Schoinos & 12 & Schoinos \\ \hline
HAC single & 2 & 32 & Khorsia & 32 & Khorsia \\ \hline
HAC single & 1 & 43 & Pagai & 43 & Pagai \\ \hline
HAC single & 3 & 45 & Megara & 45 & Megara \\ \hline
HAC single & 26 & 57 & Koropi & 57 & Koropi \\ \hline
HAC single & 1 & 77 & Krommyon & 77 & Krommyon \\ \hline
HAC single & 34 & 96 & Berbati & 96 & Berbati \\ \hline
HAC single & 1 & 109 & Sikyon & 109 & Sikyon
\end{tabular}
\end{center}
\caption{Results of partitioning the data into eight clusters and then finding the geometric centre. The normal distance data set was use to define the distances between sites and then standard clustering methods were applied: k-means or various versions of hierarchical agglomerative clustering (HAC).  Once these clusters were defined, the geometric centre of each cluster was found by finding the site closest to the mean and closest to the  median of the coordinates of the sites in each cluster. The results for one cluster are shown on each line.}
\label{tclustercentre}
\end{table}

%% **********************************************************************
%\section{Additional Material}
%
%Slides from a talk based on this work can be found at
%\\{}
%\href{http://dx.doi.org/10.6084/m9.figshare.3840249}{\texttt{TimEvansKonstanz2016.pptx}}.

%% ************************************************************************
%\section{Manuscript Formatting}
%
%\subsection{Figures}
%Frontiers requires figures to be submitted individually, in the same order as they are referred to in the manuscript. Figures will then be automatically embedded at the bottom of the submitted manuscript. Kindly ensure that each table and figure is mentioned in the text and in numerical order. Figures must be of sufficient resolution for publication \href{http://home.frontiersin.org/about/author-guidelines}{see Resolution Requirements for examples and minimum requirements}. Figures which are not according to the guidelines will cause substantial delay during the production process. Please see \href{http://home.frontiersin.org/about/author-guidelines}{General Style Guidelines for Figures in here} for full Figure guidelines
%
%
%\subsection{Tables}
%Tables should be inserted at the end of the manuscript. Please build your table directly in LaTeX.Tables provided as jpeg/tiff files will not be accepted. Please note that very large tables (covering several
%pages) cannot be included in the final PDF for reasons of space. These tables will be published as \href{http://home.frontiersin.org/about/author-guidelines}{Supplementary Material} on the online article
%page at the time of acceptance. The author will be notified during the typesetting of the final article if this is the case.
%

\end{document}